\documentclass[10pt,twocolumn,english,aps,manuscript]{revtex4}
\usepackage[T1]{fontenc}
\usepackage[latin9]{inputenc}
\usepackage{color}
\usepackage{array}
\usepackage{verbatim}
\usepackage{amsmath}
\usepackage{amssymb}
\usepackage{graphicx}

\makeatletter

\providecommand{\tabularnewline}{\\}

\@ifundefined{textcolor}{}
{%
 \definecolor{BLACK}{gray}{0}
 \definecolor{WHITE}{gray}{1}
 \definecolor{RED}{rgb}{1,0,0}
 \definecolor{GREEN}{rgb}{0,1,0}
 \definecolor{BLUE}{rgb}{0,0,1}
 \definecolor{CYAN}{cmyk}{1,0,0,0}
 \definecolor{MAGENTA}{cmyk}{0,1,0,0}
 \definecolor{YELLOW}{cmyk}{0,0,1,0}
}

\makeatother

\usepackage{babel}
\begin{document}

\title{Non-Abelian Topological Order on the Surface of a 3D Topological
Superconductor from an Exactly Solved Model}

\author{Lukasz Fidkowski$^{1}$, Xie Chen$^{2}$, Ashvin Vishwanath$^{2}$}

\affiliation{1. Department of Physics and Astronomy, Stony Brook University, Stony
Brook, NY 11794-3800, USA.}

\affiliation{2. Department of Physics, University of California, Berkeley, California
94720, USA}
\begin{abstract}
Three dimensional topological superconductors (TScs) protected by
time reversal ($\mathcal{T}$) symmetry are characterized by gapless
Majorana cones on their surface. Free fermion phases with this symmetry
(class DIII) are indexed by an integer $\nu$, of which $\nu=1$ is
realized by the B-phase of superfluid $^{3}He$. Previously it was
believed that the surface must be gapless unless time reversal symmetry
is broken. Here we argue that a fully symmetric and gapped surface
is possible in the presence of strong interactions, if a special type
of topological order appears on the surface. The topological order
realizes $\mathcal{T}$ symmetry in an anomalous way, one that is
impossible to achieve in purely two dimensions. For {\normalsize odd
$\nu$} TScs, the surface topological order must be non-Abelian. We
propose the simplest non-Abelian topological order that contains electron
like excitations, $SO(3)_{6}$, with four quasiparticles, as a candidate
surface state. Remarkably, this theory has a hidden $\mathcal{T}$
invariance which however is broken in any 2D realization. By explicitly
constructing an exactly soluble Walker-Wang model we show that it
can be realized at the surface of a short ranged entangled 3D fermionic
phase protected by $\mathcal{T}$ symmetry, with bulk electrons trasforming
as Kramers pairs, i.e. ${\cal T}^{2}=-1$ undert time reversal. We
also propose an Abelian theory, the semion-fermion topological order,
to realize an even $\nu$ TSc surface, for which an explicit model
is derived using a coupled layer construction. We argue that this
is related to the $\nu=2$ TSc, and use this to build candidate surface
topological orders for $\nu=4$ and $\nu=8$ TScs. The latter is equivalent
to the three fermion state which is the surface topological order
of a $\mathbb{Z}_{2}$ bosonic topological phase protected by $\mathcal{T}$
invariance. One particular consequence of this is that a $\nu=16$
TSc admits a trivially gapped ${\cal T}$-symmetric surface.

\begin{comment}
Non Abelian Topological Order on the Surface of a 3D Topological Supercoductor
from an Exactly Soluble Model
\end{comment}

\end{abstract}
\maketitle

\section{Introduction}

Recently it was pointed out that there exist exotic varieties of insulators
and superconductors which form distinct phases of matter. This distinction
is based on topological properties and hence falls outside the Ginzburg-Landau-Wilson
symmetry based classification. On the other hand, these distinctions
are often present only in the presence of certain symmetries, (e.g.
time reversal ${\cal T}$), leading to the terminology ``symmetry
protected topological phases'' (SPTs). Many such phases can be realized
at the level of non-interacting fermions \citep{Hasan2010,Qi2011,Hasan2011}
and several experimental realizations now exist. Their hallmark signature
is the existence of gapless edge or surface modes. The best known
example is the three dimensional topological insulator, which is protected
by charge conservation and time reversal symmetry and has a single
Dirac cone on its two dimensional surface. Another example is the
three dimensional topological superconductor. Here, time reversal
symmetry protects gapless Majorana cones at the surface. Within a
free fermion (quadratic Hamiltonian) description, different topological
phases are labeled by an integer $\mathcal{\nu}$. The gaplessness
of these surfaces is protected by the symmetry.

While a fairly complete picture exists of free fermion topological
phases \citep{Ludwig,Kitaev}, less is known about topological phases
in the presence of interactions. Even if we restrict attention to
SPTs (which may be adiabatically connected to a trivial gapped phase
in the absence of symmetry), the qualitatively new phenomena that
occur when strong interactions are present are only now beginning
to be studied. Two advances in this general area include the result
that the free fermion classification of SPT phases may be reduced
in the presence of interactions, as shown for a class of 1D topological
superconductors\citep{Chen1d,Fidkowski1d,Turner1d,Tang2012} and a
class of 2D topological superconductors with $\mathbb{Z}_{2}$ or
reflection symmetry \citep{Qi2012,Ryu2012,Yao2012,Gu2013}, and the
discovery of SPT phases of \emph{bosons} in different dimensions,
that necessarily require interactions for their realization\citep{Chen2012,Kitaev,Chen2d,LevinGu,Lu2012a,VandS,Xu2012,Max2013}.

An important lesson that emerged from studying interacting bosonic
SPT phases is that the surface of a 3D phase can be gapped \emph{without}
breaking the symmetry (either spontaneously or explicitly) if the
surface develops topological order. In other words, the system must
have deconfined anyons at the gapped, symmetric surface. Moreover,
the anyons transform under the symmetries in a way that is disallowed
in a strictly two dimensional system\citep{VandS}. Explicit examples
of such symmetric topologically ordered surfaces have been constructed
for 3D bosonic SPT phases \citep{WangSenthil2013,Max2013,bosWW}.
The simplest example is a bosonic version of the 3D topological superconductor
protected by time reversal symmetry. In a conventional surface termination,
time reversal symmetry is broken at the surface, leading to a gap.
Domain walls between time reversed domains carry chiral edge modes
with net chiral edge central charge $c_{-}=8$, and so each domain
is associated with half that chiral central charge. However it was
realized that the surface can be gapped while retaining time reversal
symmetry if it acquires the 3-fermion \citep{Kitaev} $\mathbb{Z}_{2}$
gauge theory topological order\citep{VandS}. If this topological
order were realized in 2D, it would always break $\mathcal{T}$ symmetry,
since it is associated with chiral edge states with $c_{-}=4$ (exactly
half the value associated with the surface domain walls) . However,
when realized on the boundary of a 3D system, no edge is present,
and the theory can remain time reversal symmetric. Explicit constructions
of such surface phases were given in \citep{WangSenthil2013,bosWW}. 

While 3D SPT phases of bosons remain to be experimentally realized,
it is interesting to consider the analogous topologically ordered
surface states for the 3D fermionic topological insulators and superconductors,
which we know are realized in Nature. The topologically ordered surface
states provide a rare example of a qualitatively new phenomenon that
lies beyond the free fermion description of these phases. Moreover,
as we will see shortly, we will require non-Abelian topological order
in some cases. This is a rare example where preserving symmetry requires
not just topological order, but also particles with non-abelian statistics,
and may eventually help in realizing these exotic excitations. 

In this paper we propose gapped, topologically ordered, ${\cal T}$-invariant
terminations for 3D topological superconductors (topological insulators
will be discussed in a seperate publication). In terms of the free
fermion classification these correspond to class DIII, where one has
a $\mathbb{Z}$ classification \citep{Ludwig,kitaev_periodic_table};
the $\nu=1$ member of this class is the Balian-Werthamer (BW) \citep{BW,volovik}
state of the $B$ phase of liquid $^{3}He$, whose single Majorana
cone describing the surface dispersion is protected by time reversal
in the non-interacting setting. A principal feature of this surface
is that a domain wall between regions of opposite ${\cal T}$ breaking
is associated with a single chiral Majorana mode, which has chiral
central charge $c_{-}=\frac{1}{2}$. Hence each domain may be associated
to $c_{-}=\frac{1}{4}$, something that is impossible in a purely
2d fermionic system without topological order, where $c_{-}$ is quantized
in units of $\frac{1}{2}$. One strategy is then to look for patterns
of fermionic 2d topological order (i.e. those that contain a fundamental
fermion that has trivial braiding statistics with every other excitation
and can be identified with the electron), and attempt to find a theory
with both $c_{-}=\frac{1}{4}$ mod $\frac{1}{2}$ and ${\cal T}$
symmetry. Fortunately this approach turns out to be fruitful, because
2d fermionic topological orders are extremely constrained.

First, we can rule out abelian theories, for the following reason:
the fusion/braiding universality class of any fermionic abelian topological
order can be realized in a $K$-matrix formulation, which has integral
chiral central charge. Adding in layers of $p+ip$ superconductor
- which does not change the anyon spectrum - can only change this
chiral central charge in integer multiples of $\frac{1}{2}$, and
can never generate $c_{-}=\frac{1}{4}$ mod $\frac{1}{2}$. A well
known non-Abelian example is the Moore-Read Pfaffian state \citep{MooreRead},
but it is neither ${\cal T}$-invariant nor has the correct $c_{-}$
\footnote{Although a related theory, where one reverses the direction of the
Majorana mode, can be made ${\cal T}$ invariant. Since this may have
application to 3D topological insulators, it will be discussed in
a separate publication.
}. However, the Pfaffian, with $12$ quasiparticle types, is not the
simplest non-Abelian fermionic theory. Indeed, from a classification
standpoint, the smallest such theory contains only $4$ quasiparticle
types: in addition to the trivial particle and the electron, there
is a self-semion $s$ and its time reversed partner $\tilde{s}$.
This is the integral spin sub-theory of $SU(2)_{6}$, equivalent to
${}SO(3)_{6}$ \citep{Wang,Parsa}. Serendipitously, $SU(2)_{6}$
has $c_{-}=\frac{9}{4}=\frac{1}{4}$ mod $\frac{1}{2}$, and the braiding
and fusion rules of its integral spin sub-theory are invariant under
a ${\cal T}$ symmetry which exchanges $s$ and $\tilde{s}$. A strict
2d realization of this phase should break ${\cal T}$ as we argue
below, based on the nontrivial edge chiral central charge $c_{-}=\frac{9}{4}$
in one realization. However, it is conceivable that it may appear
as a ${\cal T}$ invariant surface state of a 3D bulk, based on the
statistics of the excitations. In that case one cannot interrogate
the edge content, and hence a hidden ${\cal T}$ symmetry may exist.
We substantiate this claim by explicitly constructing a 3D ${\cal T}$
symmetric model whose surface displays the ${}SO(3)_{6}$ topological
order. 

Our central tool is the Walker-Wang (WW) construction \citep{WW,Burnell},
which, in essence, converts a given surface topological order into
a prescription for the bulk wave function (similar to the connection,
one dimension lower, between quantum Hall wave functions and edge
conformal field theories). The construction also provides an exactly
soluble model to realize this wave function and surface topological
order - so we know explicitly that it can be realized at the surface
of an appropriate 3D system. Moreover in the case of fermionic topological
phases it allows us to fix the transformation law for fermions under
the protecting symmetry. Here we will see that the electrons must
transform projectively, i.e. with ${\cal T}^{2}=-1$ under time reversal
symmetry. For constructing the exactly soluble 3D models, one is given
the surface topological order, specified in terms of quasiparticle
labels and fusion and braiding rules, collectively denoted $\mathcal{B}$
(mathematically, a premodular unitary category). When the only particle
that braids trivially with everything in $\mathcal{B}$ is the identity
- the so-called modular case - the WW models realize a confined 3+1
dimensional phase, with topological order $\mathcal{B}$ on the surface.
Although the bulk is trivial, these models sometimes naturally allow
for an incorporation of symmetry that can protect the surface topological
order and hence result in a non-trivial 3D bosonic SPT \citep{bosWW}.
This strategy was successfully applied to construct WW models that
realize the 3 fermion surface topological order with time reversal
symmetry, which realizes a 3D bosonic SPT phase \citep{bosWW}. In
the present case, however, the input $\mathcal{B}=SO(3)_{6}$ contains
the electron, which has trivial braiding statistics with all other
excitations. We will see that this leads to a 3D WW model whose only
bulk deconfined excitation is a ${\cal T}^{2}=-1$ fermion - the electron
- and whose surface realizes ${}SO(3)_{6}$. We note a minor caveat
here. Since it is convenient to work entirely in terms of a bosonic
WW model, rather than introducing fundamental fermions (electrons),
the deconfined bulk fermions carry $\mathbb{Z}_{2}$ gauge charge,
and realize the $\mathbb{Z}_{2}$ gauged 3D topological superconductor.
Ungauging this theory by suppressing $\mathbb{Z}_{2}$ flux loops
is straightforward and yields the topological superconductor. 

Finally, we discuss Abelian topological orders which are candidates
for surfaces of even $\nu$ topological superconductors. Specifically,
we propose three such topological orders - the semion-fermion model,
the doubled semion-fermion model, and the fermionization of the bosonic
SPT. The semion-fermion model is simply $U(1)_{2}$ times $\{1,f\}$
where $f$ is the electron; however, time reversal acts in a non-trivial
way, exchanging the semion $s$ with $sf$. We argue that this theory
cannot be realized in 2D with ${\cal T}$ symmetry, but can appear
on the surface of a 3D topological phase. In addition to a Walker-Wang
model, we provide a coupled layer construction of this phase. An important
point is that there are two varieties of the semion-fermion model,
distinguished by the action of ${\cal T}^{2}$. These two are opposite
to each other in the sense that placing one on top of the other allows
one to condense everything and yields a trivial ${\cal T}$-symmetric
theory, i.e. one corresponds to $\nu$ and the other to $-\nu$. However,
two copies of the same variety yield the double semion-fermion model.
Two copies of the latter yield the three fermion $\mathbb{Z}_{2}$
gauge theory discussed above in the context of the bosonic SPT, times
$\{1,f\}$. We argue that this fermionic theory cannot be realized
in $2D$ with ${\cal T}$ symmetry either, but two copies of it can
be. It has been conjectured previously (\citet{Kunpublished}) that
the non-interacting $\mathbb{Z}$ classification of 3D topological
superconductors breaks down to $\mathbb{Z}_{16}$; if the $SO(3)_{6}$
and semion-fermion topological superconductors we construct adiabatically
connect to free fermion topological superconductors, then the three
Abelian topological orders must correspond to $\nu=2$ mod $4$, $\nu=4$
mod $8$, and $\nu=8$ mod 16. Taken together, these would then give
topological terminations for all free fermion topological superconductors.

This paper is organized as follows. In Section\ref{sec:SO} we treat
the $SO(3)_{6}$ topological order, corresponding to $\nu=1\,{\rm mod}\ 2$.
An intuitive discussion of why it can be realized at the surface of
a 3D ${\cal T}$ SPT of electrons is presented in subsection \ref{sec:Time-Reversal-Symmetry},
and the Walker Wang is discussed in the following two subsections.
Next, in Section \ref{sec:SemionFermion}, the abelian topological
orders are discussed, starting with the semion-fermion state, for
which we also provide a coupled layer construction. We provide an
argument for the semion-fermion state being realized at the surface
of a $\nu=2$ mod $4$ TSc, and then derive the abelian topological
orders corresponding to $\nu=4$ mod $8$, and $\nu=8$ mod 16. The
Conclusions connect our results with the classification of fermionic
TScs and comment on future directions; the appendices contain a construction
of a 2D $\mathbb{Z}_{2}$ gauge theory in which ${\cal T}$exchanges
the $\mathbb{Z}_{2}$charge $e$ and the $\mathbb{Z}_{2}$flux $m$
particles, and a discussion of the stability of bosonic SPT phases
protected by ${\cal T}$, in the presence of fundamental electrons.

\section{$SO(3)_{6}$ Topological Order and Walker-Wang Construction of the
3D Phase\label{sec:SO}}

\subsection{\textmd{$SO(3)_{6}$} Topological Order of Surface State\label{sub:NonabelianTO}}

\begin{figure*}
\includegraphics[bb=0bp 0bp 1024bp 768bp,clip,scale=0.35]{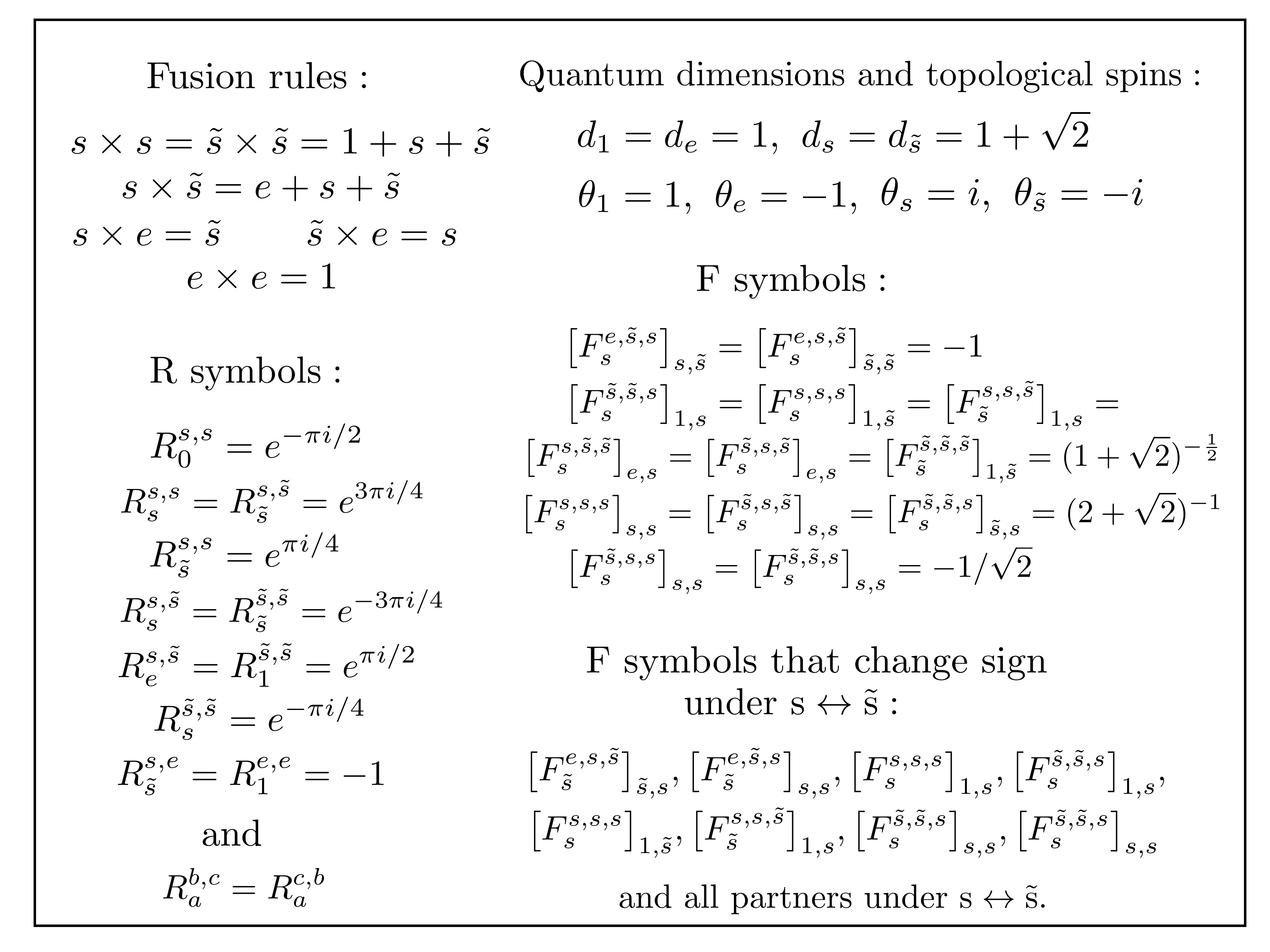}\caption{\label{fig:collected_data}Braiding statistics and fusion data for
$SO(3)_{6}$.}
\end{figure*}

Before delving into the construction of the Walker-Wang model, let
us describe the fusion and braiding properties of ${}SO(3)_{6}$.
A useful viewpoint on this phase is to begin with the well known topological
order SU(2)$_{6}$ \citep{Witten}, which is a bosonic Read-Rezayi
state \citep{Parsa,ReadRezayi} with six quasiparticles labeled by
spins $j\in\{0,\,\frac{1}{2},\,1,\,\frac{3}{2},\,2,\,\frac{5}{2},\,3\}$,
and is described by the SU(2)$_{k=6}$ Wess-Zumino-Witten chiral edge
theory with chiral central charge $c_{-}=\frac{9}{4}$. The topological
spins of the quasiparticles are $\theta_{j}=e^{i2\pi\frac{j(j+1)}{8}}$.
Thus $j=1$ and $j=2$ particles are self-semions, with topological
spins $\pm i$, while the $j=3$ particle is a fermion. However, the
fermion has mutual statistics with the half-integer spin particles.
The latter can be eliminated if we introduce fundamental fermions
(electrons) into the theory and condense the bound state of the $j=3$
particle and the electron. This bound state can be condensed since
it is a self-boson. However, since it has mutual statistics with the
half-integer spin particles $j=\frac{1}{2},\,\frac{3}{2},\,\frac{5}{2}$,
they are confined, and one is left with just the integer spin particles.
Of these, the fourth particle $j=3$ has trivial braiding statistics
with the remaining excitations, and is identified as the electron
(note that this results in a non-modular theory, as in any topological
order that contains fundamental fermions). Since only the integer
representations of the spins remain in the final theory, we call it
SO(3)$_{6}$ topological order 
\footnote{Sometimes, only theories with $k=0\,{\rm mod\,4}$ are labeled SO(3)$_{k}$,
since otherwise they are non-modular. Since we are specifically interested
in theories which contain the electron, which are necessarily non-modular,
we will not make this distinction.}. The condensation process that converts the SU(2)$_{6}$ to SO(3)$_{6}$
does not change the edge central charge and hence the final theory
is also expected to have $c_{-}=\frac{9}{4}$. We henceforth denote
the two self-semions by $s,\tilde{s}$, and the fermion by $e$: $\{0,1,2,3\}\rightarrow{\{1,s,\tilde{s},e\}}$.
Their fusion rules and topological spins are shown in Fig. \ref{fig:collected_data}.

\begin{figure}
\includegraphics[bb=0bp 200bp 1200bp 760bp,clip,scale=0.35]{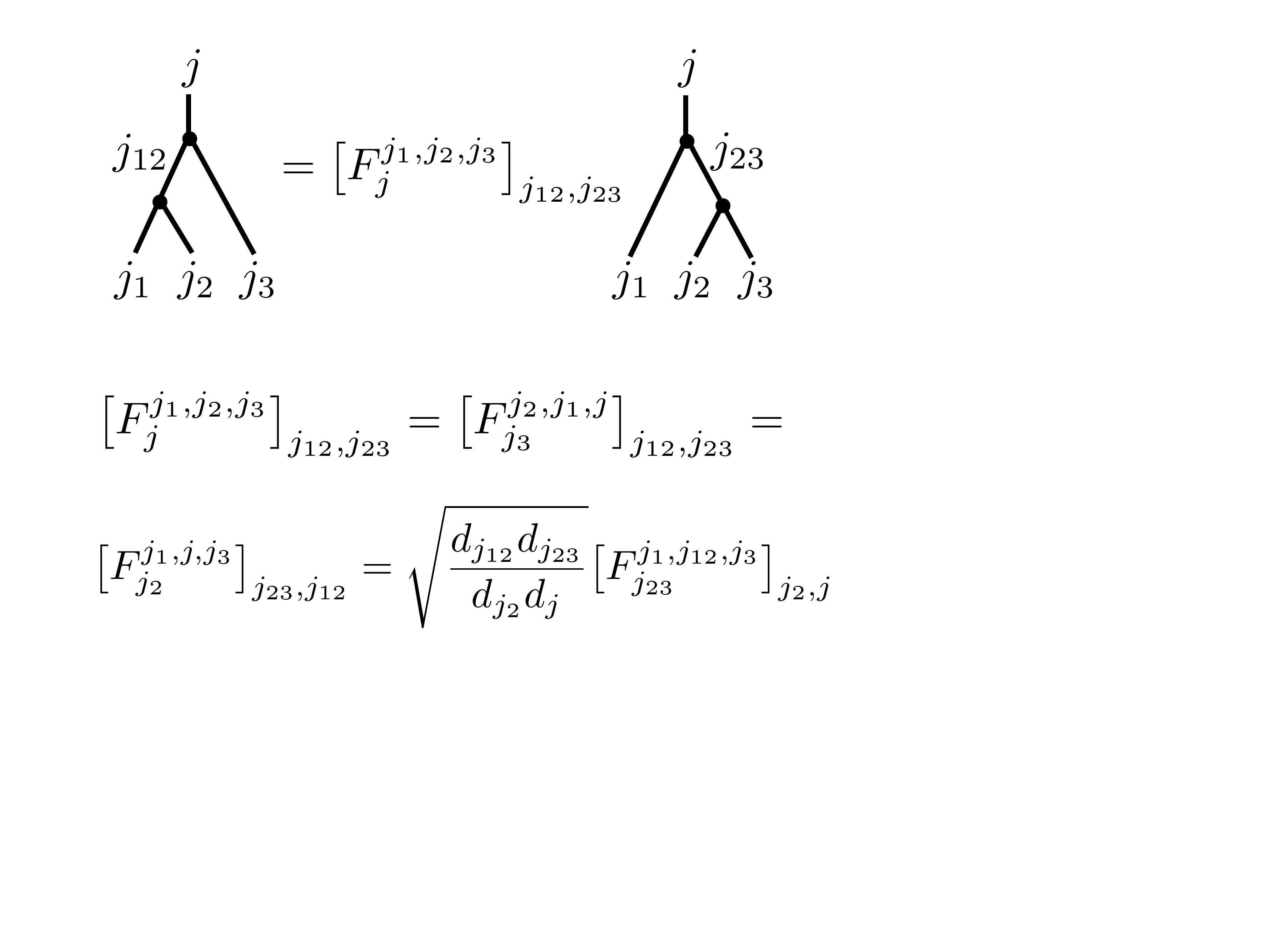}\caption{\label{fig:general_f_matrix}Graphical definition of $F$ symbol,
together with some identities satisfied by $F$.}
\end{figure}

In our analysis we will need more information about the fusion and
braiding, however: the data that enters the WW Hamiltonian requires
the $F$ and $R$ symbols, which describe associativity of fusion
(a quantum analogue of Clebsch-Gordon coefficients) and exchange of
an arbitrary pair of quasiparticles, respectively, and uniquely determine
${}SO(3)_{6}$ as a premodular category \citep{Parsa}. For the definition
of the $F$ symbols, see Fig. \ref{fig:general_f_matrix}, which also
contains some of their symmetries. All of the non-trivial $F$ symbols
(i.e. those not equal to $1$) can be obtained from those in Fig.
\ref{fig:collected_data} by using compositions of these symmetries,
together with their invariance up to sign under $s\leftrightarrow\tilde{s}$.
A representative set of $F$-symbols which change sign under this
exchange are also shown in Fig. \ref{fig:collected_data}.

\subsection{Time Reversal Symmetry and \textmd{$SO(3)_{6}$} Topological Order
\label{sec:Time-Reversal-Symmetry}}

We first elaborate on the question of time reversal symmetry and the
$SO(3)_{6}$ topological order. As discussed in the overview, since
this topological order is obtained by condensing particles in the
$SU(2)_{6}$ theory, which is modular and has an edge central charge
$c_{-}=\frac{9}{4}$, we expect the same edge central charge in at
least some realizations of this phase. Could one potentially realize
the same $SO(3)_{6}$ topological order with a trivial edge? We now
argue that it is impossible to realize this topological order in 2D
without a chiral edge mode. This guarantees that it cannot be realized
in a 2D system with $\mathcal{T}$ symmetry, and thus, if realized
on the surface of a 3D $\mathcal{T}$ symmetric system, defines a
3D topological phase. For modular topological orders, where all non-trivial
quasiparticles have non-trivial braiding statistics with at least
one other particle, a powerful formula relates the topological properties
to the chiral edge central charge $c_{-}\,{\rm mod}\,8$ \citep{Kitaev}:

\begin{equation}
\frac{1}{\mathcal{D}}\sum_{a}{d_{a}^{2}}\theta_{a}=e^{i2\pi c_{-}/8}\label{eq:edge-c}
\end{equation}

\noindent where $d_{a}$ ($\mathcal{D}$) are the individual (total)
quantum dimensions of the quasiparticles. Unfortunately, we cannot
directly apply this formula since we have a non-modular theory (the
electron has trivial braiding with all quasiparticles). In fact we
can easily see that the edge central charge can be changed by units
of $c_{-}=\frac{1}{2}$ without affecting the topological order, simply
by putting the electrons in a $p\pm ip$ topological superconductor
phase. %
\begin{comment}
\noindent - this generates $16$ theories, all with nonzero chiral
central charge \citep{KitaevHoneycomb}. Are there other modular extensions,
potentially with $c_{-}=0$?
\end{comment}
{} This suggests that the edge of $SO(3)_{6}$ is necessarily chiral.
This result can be further argued as follows. Imagine that there was
also a realization of the same topological order with a $c_{-}=0$
edge, and consider it in conjunction with the existing $c_{-}=\frac{9}{4}$
realization. Then, we could perform a reflection on the first phase
and combine it with the second. In this way we will have realized
a quantum double model, which can always be confined by condensing
a bosonic quasiparticle. In this process the edge central charge will
not change and will remain $c_{-}=\frac{9}{4}$. However, we now have
eliminated all topological order. A fermionic system with no topological
order must have half integer quantized edge central charge, so we
have arrived at a contradiction. Therefore there cannot be a 2D system
with $SO(3)_{6}$ topological order and without a chiral edge state.

Let us now consider time reversal purely at the level of fusion and
braiding rules. An examination of the topological spins in our theory
(Fig. \ref{fig:collected_data}) suggests that it is time reversal
invariant under the exchange $s\leftrightarrow\tilde{s}$. Indeed,
such an exchange together with complex conjugation gives a set of
$F$ and $R$ symbols which must be gauge equivalent to the original
ones, because ${}SO(3)_{6}$ is the unique theory with these topological
spins and fusion rules. However, ${\cal T}$-invariance of the WW
Hamiltonian (constructed in the next section) requires more: we will
need a time reversal transformation law which leaves the $F$ and
$R$ symbols exactly invariant, not just invariant up to gauge transformation.
Quick examination of $\big[F_{e}^{s,s,s}\big]_{\tilde{s},\tilde{s}}$
and $\big[F_{e}^{\tilde{s},\tilde{s},\tilde{s}}\big]_{s,s}$ shows
that there is no gauge in which such a transformation law takes the
simple form $s\leftrightarrow\tilde{s}$ followed by complex conjugation
- this is because these two $F$ symbols differ by a sign in our gauge
and transform by complex conjugate phases under a gauge transform,
so they cannot be complex conjugates in any gauge. However, the following
more general ${\cal T}$ transformation law does work, and in fact
forces ${\cal T}^{2}=-1$ on the electrons: we define ${\cal T}$
as the operation that exchanges $s\leftrightarrow\tilde{s}$, complex
conjugates, and multiplies by certain phase factors $\alpha_{c}^{a,b}$
(see below) associated to the fusion spaces $V_{c}^{a,b}$ in our
anyon theory
\footnote{This structure is reminiscent of the notion of $G$ action in a braided
$G$-crossed category \citep{Etingof,DGNO}
}. It is this ${\cal T}$ transformation which commutes with all of
the $F$ and $R$ symbols; link invariants (i.e. braiding amplitudes)
computed in the corresponding picture calculus are then invariant
under it.

The phase factors $\alpha_{c}^{a,b}$ are defined as follows. For
vertices of type $(s,\tilde{s},e)$, $\alpha_{c}^{a,b}$ is $\pm i$
depending on the sign of the permutation that takes $(s,\tilde{s},e)$
to $(a,b,c)$. Specifically,

\begin{alignat*}{1}
\alpha_{e}^{s,\tilde{s}} & =\alpha_{s}^{\tilde{s},e}=\alpha_{\tilde{s}}^{e,s}=i,\\
\alpha_{\tilde{s}}^{s,e} & =\alpha_{s}^{e,\tilde{s}}=\alpha_{e}^{\tilde{s},s}=-i.
\end{alignat*}

\noindent For vertices at which one or more of $a,b,c$ is the trivial
anyon $1$, we set $\alpha_{c}^{a,b}=1$. The remaining vertices have
all $a,b,c$ either $s$ or $\tilde{s}$. We take

\begin{alignat*}{1}
\alpha_{s}^{s,s}=\alpha_{\tilde{s}}^{\tilde{s},\tilde{s}}=i
\end{alignat*}

\noindent and $\alpha_{c}^{a,b}=-i$ when $(a,b,c)$ consists of some
permutation of two $s$'s and one $\tilde{s}$ or vice versa.

With this definition, ${\cal T}$ commutes with $F$. Since in our
gauge the $F$ symbols are all real, this reads:

\begin{equation}
\alpha_{n}^{j,k}\alpha_{l}^{i,n}\big[F_{l}^{i,j,k}\big]_{m,n}=\big[F_{\tilde{l}}^{\tilde{i},\tilde{j},\tilde{k}}\big]_{\tilde{m},\tilde{n}}\alpha_{m}^{i,j}\alpha_{l}^{m,k}
\end{equation}

\noindent where $i\rightarrow\tilde{i}$ is just the permutation that
fixes $i=1,e$ and exchanges $i=s,\tilde{s}$. ${\cal T}$ also commutes
with $R$:

\begin{center}
\begin{equation}
\alpha_{k}^{j,i}\,(R_{k}^{i,j})^{*}=R_{\tilde{k}}^{\tilde{i},\tilde{j}}\,\alpha_{k}^{i,j}
\end{equation}

\par\end{center}

\noindent These two relations are sufficient to show that Wilson lines
computed in the corresponding picture calculus are invariant under
${\cal T}$. Note that $\big(\alpha_{\tilde{k}}^{\tilde{i},\tilde{j}}\big)\big(\alpha_{k}^{i,j}\big)^{*}=1$
for all vertex types except those where $(i,j,k)$ is a permutation
of $(s,\tilde{s},e)$, for which $\big(\alpha_{\tilde{k}}^{\tilde{i},\tilde{j}}\big)\big(\alpha_{k}^{i,j}\big)^{*}=-1$.
This means that ${\cal T}^{2}$ fixes all particle types, and gives
minus signs to $(s,\tilde{s},e)$ vertices. In any string net that
respect the fusion rules there are an even number of such vertices,
so ${\cal T}^{2}$ acts as the identity on them. In the exactly solved
Walker-Wang model of the next section, this fact translates to the
many body ground state being invariant under ${\cal T}$, with ${\cal T}^{2}=1$.
However, string-net configurations of the Walker-Wang model also include
fusion-rule violating vertices such as $(1,1,e)$; a consistent definition
of time reversal that includes these will require them to have ${\cal T}^{2}=-1$,
which will necessitate decorating the Walker-Wang model with additional
spin $1/2$'s (Kramers doublets). In the next section we will see
how to do this by binding Haldane chains to the $e$ links of the
model. Ultimately, this ensures that the fundamental fermions in the
theory (the electrons) must transform under time reversal as ${\cal T}^{2}=-1$.
Since this is a important point, we elaborate on its origin in a more
intuitive way below.

\paragraph{Emergence of $\mathcal{T}^{2}=-1$ for fermions in the Walker-Wang
model:}

\begin{figure}[htbp]
\includegraphics[width=0.9\linewidth]{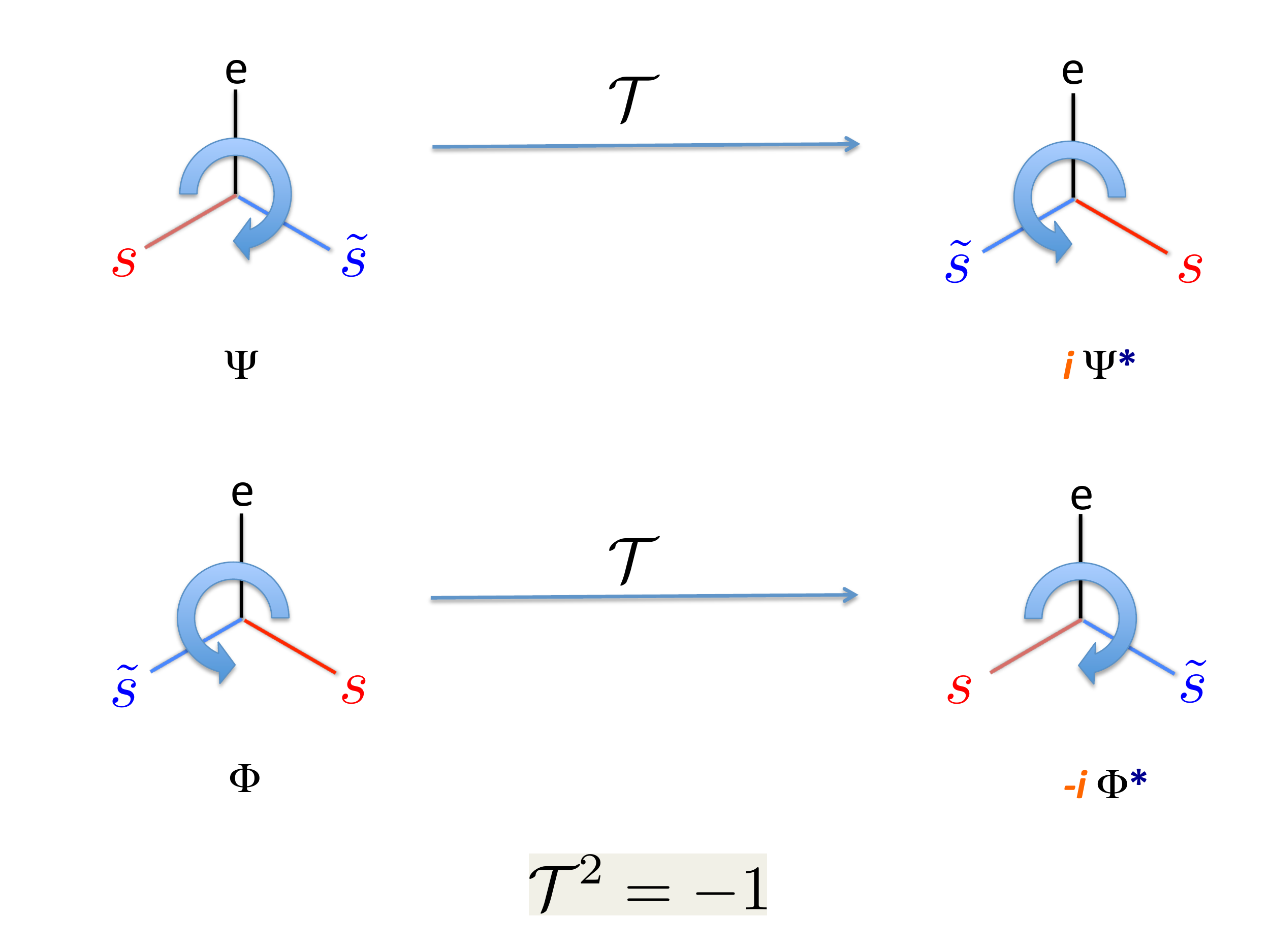} \caption{(color online) Origin of ${\mathcal{T}}^{2}=-1$ in Walker Wang models
of SO(3)$_{6}$ surface topological order. The additional phase factor
shown in red depends on the orientation of the vertices and is required
to ensure time reversal symmetry. This leads to ${\mathcal{T}}^{2}=-1$
for fermions in this model.}

\label{fig:Spin} 
\end{figure}
Let us briefly review the idea behind the Walker-Wang construction
in order to make this point. The Hilbert space, defined on the links
of the 3D lattice model, contains a state for each quasiparticle in
the theory - in this case there are four states per link, labeled
by the four particle types. The ground state wave function is a quantum
superposition of loops labeled by these four indices (or four colors).
At the vertices (we consider a trivalent lattice for simplicity -
it is always possible to deform a lattice such as a cubic lattice
into a trivalent one by splitting vertices), the loops are allowed
to branch and combine according to the fusion rules. The amplitudes
for these loop configurations are determined by imagining them as
space-time Wilson loops of the $2+1D$ TQFT representing the surface
topological order, and calculating the quantum amplitudes within that
theory. Thus: $\Psi_{{\rm 3D}}(C)=\langle W(C)\rangle_{{\rm 2+1TQFT}}$.
In practice, these amplitudes are implemented by writing down a parent
Hamiltonian. Consider a part of the ground state wave function as
shown in the top-left of Figure \ref{fig:Spin}. The $s$ and $\tilde{s}$
particles of the theory, which have semion and ``anti''-semion self
statistics, fuse to give the electron $e$. Say this has an associated
amplitude $\Psi$. Then, in order to preserve time reversal symmetry,
one needs to not only complex conjugate this amplitude, but also multiply
by a phase factor $\pm i$. Only then does the transformation commute
with the $\mathcal{R}$ moves, and produce a time reversal symmetric
wave function. The choice of phase factor is fixed by the orientation
of the vertex - eg. if we choose $i$ for the vertex where the $\{e,\tilde{s},s\}$
appear on going around anti clockwise, we must choose $-i$ for the
vertex of the opposite sense shown on the bottom left. Since time
reversal exchanges $s$ and $\tilde{s}$, and hence these two vertices,
we have the sequence $\Psi\overset{\mathcal{T}}{\rightarrow}i\Psi^{*}\overset{\mathcal{T}}{\rightarrow}-i(i\Psi^{*})^{*}=-\Psi$
and thus $\mathcal{T}^{2}=-1$. Since this vertex results in the creation
of a fermion, we will ultimately associate this transformation law
to the fermions in the theory. Note that in the Walker Wang models
clockwise/counterclockwise orientations are fixed since the projection
of the 3D lattice on the 2D plane is fixed a priori in order to define
the model. Hence, one can imagine that a crystal axis is picked in
defining the models, and the remaining sense along the axis, required
to define a spinning particle, is provided by the permutation of the
three labels.

The time reversal symmetry described above assigns a phase factor
that depends jointly on the state of three bonds that meet at a vertex.
Such a non-on-site transformation, i.e. one that does not act independently
on the physical variables that reside at links, is not a fully satisfactory
implementation of symmetry. This is readily fixed by attaching additional
link variables that are put into a `Haldane chain' phase protected
by time reversal symmetry, along the $e$ strings. When these strings
end (for example by splitting into an $s,\,\tilde{s}$ pair, as in
the figure \ref{fig:Spin} ), the spin 1/2 excitation generated at
the ends automatically provides the requisite phase factors to keep
the state invariant under time reversal. Furthermore, the vertex term
violating configurations mentioned previously are also made time reversal
symmetric by this construction.

\paragraph{Variants of ${\cal T}$action on $SO(3)_{6}$:}

Another important point is that if $SO(3)_{6}$ indeed corresponds
to some odd $\nu$ topological superconductor, then one would expect
its parity conjugate - i.e. a flipped version of $SO(3)_{6}$ - to
be distinct, and correspond to $-\nu$. We address this issue for
the (easier) Abelian semion-fermion theory in the following section,
but leave the case of $SO(3)_{6}$ to future work.

\subsection{Review of Standard Walker-Wang construction\label{sec:Review-of-Standard-WW}}

\begin{figure}
\includegraphics[bb=0bp 170bp 1200bp 600bp,clip,scale=0.35]{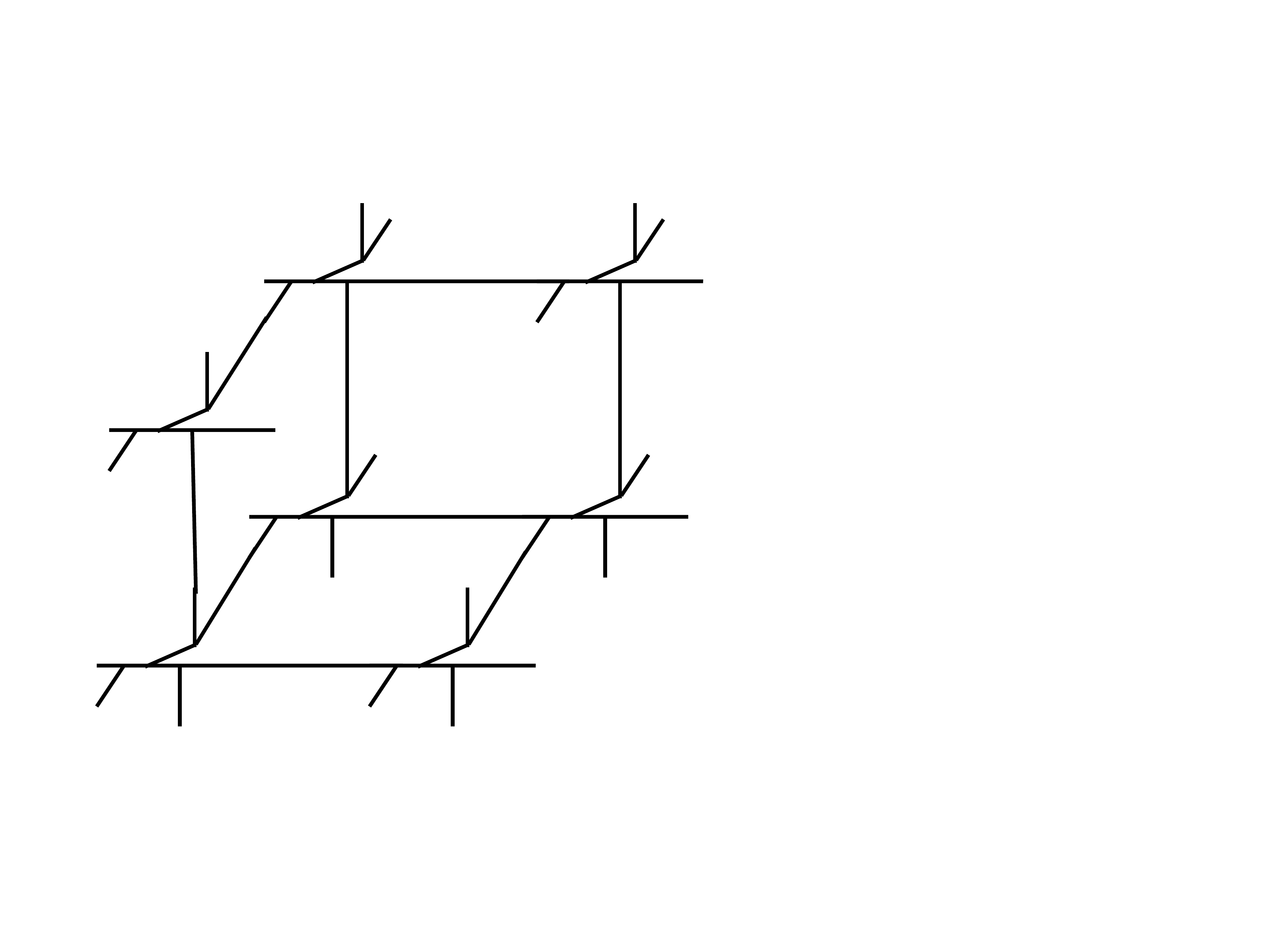}\caption{\label{fig:trivalent_lattice}Trivalent resolution of the cubic lattice
\citep{WW}.}
\end{figure}

We now construct our exactly soluble model. First we review the standard
Walker-Wang construction \citep{WW,Burnell}, for the specific case
of the premodular category $\mathcal{B}={}SO(3)_{6}$. This model
will have ${}SO(3)_{6}$ topological order on the surface and its
only deconfined bulk excitation will be a fermion. Then we describe
how to extend it by ``gluing Haldane chains''\citep{CLV} to the
$e$-lines; the resulting model will be ${\cal T}$-invariant under
a natural on-site ${\cal T}$symmetry, with the bulk deconfined fermion
- the electron - carrying ${\cal T}^{2}=-1$.

\begin{figure}
\includegraphics[bb=0bp 270bp 1200bp 600bp,clip,scale=0.35]{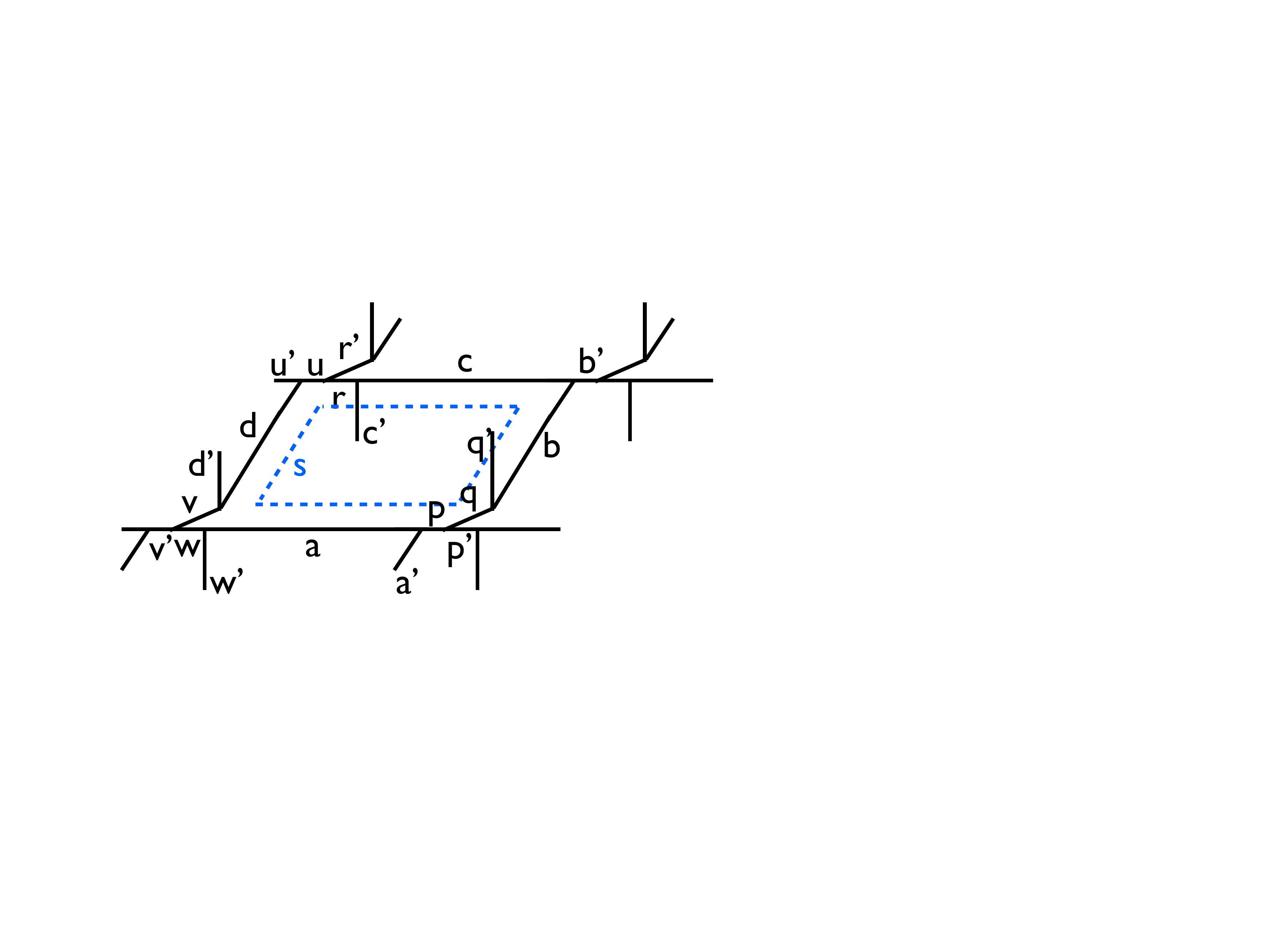}\caption{\label{fig:plaquette_term}Plaquette term in Walker-Wang model (taken
from \citep{WW}).}
\end{figure}

Informally, the states in the WW model are string-nets obeying the
fusion rules of $\mathcal{B}$. The Hamiltonian is engineered in such
a way that the ground state consists of a superposition of such string
nets with amplitudes equal to their evaluation in the picture calculus
of $\mathcal{B}$. It is important to note that the WW models work
with a particular planar projection of the 3D lattice, yielding a
natural choice of framing. The only deconfined strings correspond
to the so-called symmetric center $\mathcal{Z}(\mathcal{B})$: quasiparticles
in $\mathcal{B}$ which have trivial braiding with all other quasiparticles.
In our case, these are just the electrons. Furthermore, the statistics
of the bulk deconfined excitations are just those of $\mathcal{Z}(\mathcal{B})$,
so our bulk deconfined quasiparticle is indeed a fermion \citep{WW}
\footnote{It is actually known that the symmetric center $\mathcal{Z}(\mathcal{C})$
of any braided fusion category $\mathcal{C}$ comes in two types \citep{DR}:
either it consists entirely of bosons and is isomorphic to the set
of representations of some finite group $G$, in which case the bulk
forms a (possibly twisted) $G$-gauge theory, or a supersymmetric
version of this, where $G$ contains some odd elements and the corresponding
representations have even and odd sectors, corresponding to bosons
and fermions respectively.
}.

In order to explicitly describe the Walker-Wang model, we closely
follow \citep{WW} and refer the reader there for further details.
We start with a planar projection of a trivalent resolution of the
cubic lattice, as in Fig. \ref{fig:trivalent_lattice}. The links
are labeled with the quasiparticle types of ${}SO(3)_{6}$. The Hamiltonian
enforces fusion rules at the vertices, and contains some plaquette
terms. For any quasiparticle $s$ and plaquette $P$ (Fig. \ref{fig:plaquette_term}),
there is a term in the Hamiltonian which acts on the links of that
plaquette, labeled in Fig. \ref{fig:plaquette_term}. This term can
change the labels of these links, and can also depend on the labels
of adjoining links, which have primes on them (but cannot change these).
Explicitly, the matrix element between a state with plaquette links
$(a\, b\, c\, d\, p\, q\, r\, u\, v\, w)$ and $(a''b''c''d''p''q''r''u''v''w'')$
is

\begin{alignat}{1}
(B_{P}^{s})_{a'',\ldots,w''}^{a,\ldots,w}=\;\;\;\;\;\;\;\;\;\;\;\;\;\;\;\;\;\;\nonumber \\
R_{q}^{q'b}\big(R_{c}^{c'r}\big)^{*}\big(R_{q''}^{q'b''}\big)^{*}R_{c''}^{c'r''}\big[F_{a'}^{a'',s,p}\big]_{a,p''}\big[F_{p'}^{p'',s,q}\big]_{p,q''}\nonumber \\
\big[F_{q'}^{q'',s,b}\big]_{q,b''}\big[F_{b'}^{b'',s,c}\big]_{b,c''}\big[F_{c'}^{c'',s,r}\big]_{c,r''}\big[F_{r'}^{r'',s,u}\big]_{r,u''}\nonumber \\
\big[F_{u'}^{u'',s,d}\big]_{u,d''}\big[F_{d'}^{d'',s,v}\big]_{d,v''}\big[F_{v'}^{v'',s,w}\big]_{v,w''}\big[F_{w'}^{w'',s,a}\big]_{w,a''}.\label{eq:WW_plaq}
\end{alignat}

$ $

\noindent The intuition behind this complicated looking term is that
it fuses in the loop $s$ to the skeleton of the plaquette using multiple
$F$ moves, but in the process of doing so must use $R$ symbols to
temporarily displace certain links ($c'$ and $q'$ in Fig. \ref{fig:plaquette_term}).
The Hamiltonian then contains a sum of all these plaquette terms,
weighted by the quantum dimensions $d_{s}$. It is possible to check
that all of these terms commute, and the result is a model that satisfies
the properties described above - again we refer the interested reader
to \citep{WW} and \citep{Burnell} for more details. We have thus
constructed an exactly solved model which explicitly realizes the
$SO(3)_{6}$ theory on its surface.

\subsection{Improved Walker Wang Model - Onsite ${\cal T}$ symmetry and ungauging
the bulk topological order.\label{sub:Caveat}}

\subsubsection{Exactly soluble model with onsite ${\cal T}$-symmetry}

One way to make the model of Sec. \ref{sec:Review-of-Standard-WW}
time reversal invariant is by defining the ${\cal T}$ operator to
act by $s\leftrightarrow\tilde{s}$ on link labels and the phase factors
$\alpha_{k}^{i,j}$ on vertices. As discussed in \ref{sec:Time-Reversal-Symmetry},
this commutes with the $F$ and $R$ symbols, and hence with the Walker-Wang
Hamiltonian. However, in order to interpret this as the time reversal
invariance of the $\mathbb{Z}_{2}$-gauged topological superconductor,
we have to vary this definition slightly.

To see why, note that in the Walker Wang model, one has to make some
choice about how it will act on vertices that violate the fusion rules.
For example, consider a configuration of an electron string which
terminates at a $(1,1,e)$ fusion-rule violating vertex on one end,
and at an $(s,\tilde{s},e)$ fusion-rule respecting vertex on the
other. This is certainly not a low energy state with respect to the
vertex or plaquette terms, but it is an allowed Hilbert space configuration.
Unless we make the $(1,1,e)$ vertex into a Kramers pair, which requires
introducing extra spin-$1/2$ degrees of freedom not present in the
original Walker Wang model, we will have ${\cal T}^{2}=-1$ on this
configuration (from the one $(s,\tilde{s},e)$ vertex). But this cannot
happen in a finite $\mathbb{Z}_{2}$ gauged topological superconductor,
where the odd charge sector has been projected out. Hence we are forced
to introduce some additional spin-$1/2$ degrees of freedom in order
to have a ${\cal T}$symmetry consistent with that of a $\mathbb{Z}_{2}$-gauged
topological superconductor - this is another argument for the introduction
of additional spin $1/2$ degrees of freedom. 

\begin{figure}
\includegraphics[bb=30bp 200bp 1200bp 620bp,clip,scale=0.35]{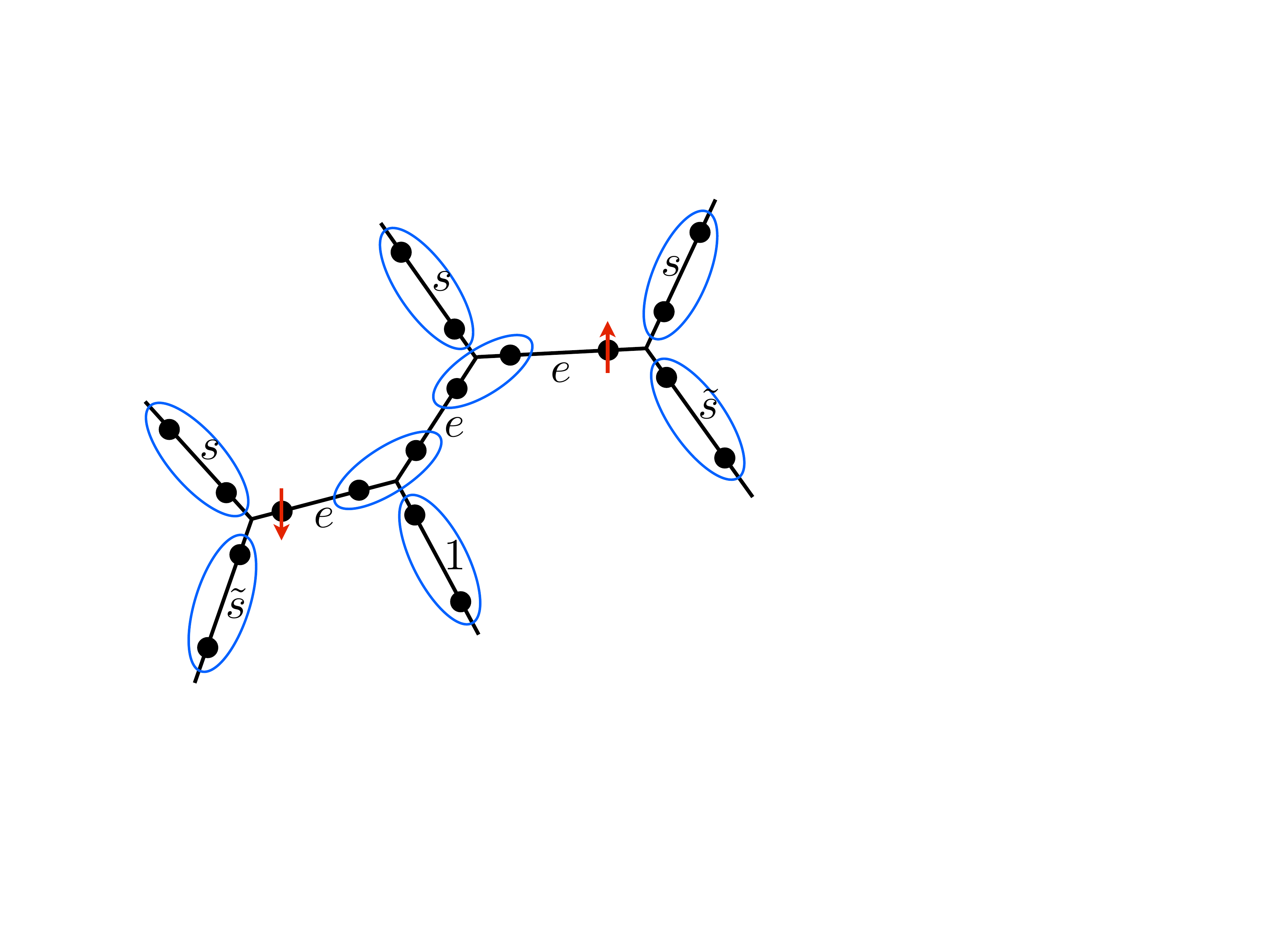}\caption{\label{fig:attaching_Haldane_chains}Sample spin configuration in
the decorated Walker-Wang model. Black dots represent spin $1/2$'s,
and blue ellipses represent spin singlets. The leftmost vertex has
a counterclockwise $(e,s,\tilde{s})$ ordering of labels, and so prefers
a down spin, whereas at the rightmost vertex the counterclockwise
label ordering is $(e,\tilde{s},s)$ and an up spin is preferred (red).}
\end{figure}

Specifically, we proceed as follows: decorate each link with four
extra states, to be thought of as two spin $1/2$'s, one associated
with each of the two vertices adjacent to the link. Then add a term
$H_{V}$ to the Hamiltonian that projects these into a singlet, unless
the link is labeled with an $e$, in which case it forces the two
spin $1/2$'s into singlets with spin $1/2$'s from other $e$ lines
adjacent to those vertices. If there are none (or if a total of $3$
$e$ lines meet at the vertex) there will be an unpaired spin $1/2$
there. This is very similar to the construction in \citep{CLV}, and
has the effect of binding Haldane chains to the $e$ lines. At a $(s,\tilde{s},e)$
vertex we also add a term to $H_{V}$ that energetically prefers either
the up spin or the down spin depending on the sign of the permutation
that takes $(s,\tilde{s},e)$ into a counter-clockwise labeling of
the three links adjoining the vertex. Finally, we modify the plaquette
terms in the original WW Hamiltonian in such a way that they move
between low energy spin configurations - for details, see Appendix
\ref{sec:Appendix-C:-Hamiltonian}.

The Hamiltonian constructed in Appendix \ref{sec:Appendix-C:-Hamiltonian}
commutes with a global symmetry ${\cal T}$ which acts on the spin
$1/2$'s in the usual way by complex conjugation followed by multiplication
by $\sigma^{y}$ (note the extra factor of $i$ in our definition
- it can be eliminated by a gauge transform if desired, it is only
${\cal T}^{2}=-1$ that is important), and on the Walker-Wang labels
by exchanging $s\leftrightarrow\tilde{s}$, and applying the phase
factors $\alpha_{c}^{a,b}$ on vertices at which all of the $a,b,c$
are either $s$ or $\tilde{s}$. Although this is not a true onsite
action, it can be made so by a suitable gauge transformation 
\footnote{Indeed, if we take a gauge transformation to act by phase factor $\beta_{c}^{a,b}$
on the fusion space $V_{c}^{a,b}$, then any choice which satisfies
$\beta_{s}^{s,s}\beta_{\tilde{s}}^{\tilde{s},\tilde{s}}=i$, $\beta_{\tilde{s}}^{s,s}\beta_{s}^{\tilde{s},\tilde{s}}=-i$,$\beta_{s}^{s,\tilde{s}}\beta_{\tilde{s}}^{\tilde{s},s}=-i$,$\beta_{s}^{\tilde{s},s}\beta_{\tilde{s}}^{s,\tilde{s}}=-i$
and sets the other $\beta_{c}^{a,b}$ to be trivial, does the trick.
The action of ${\cal T}$is then truly onsite.}. In this system, all endpoints of $e$ lines effectively contribute
a minus sign to ${\cal T}^{2}$, and since the number of such endpoints
$ $is even in any configuration, the Hilbert space has ${\cal T}^{2}=1$.

\subsubsection{Ungauging the bulk $\mathbb{Z}_{2}$ topological order.}

A second point has to do with the nature of the bulk phase in the
WW model, when the surface topological order contains an electron,
i.e. a fermion with trivial braiding statistics with all other excitations.
In this case the fermionic excitation is deconfined in the bulk. However,
note that all the degrees of freedom entering the microscopic WW model
are bosonic, since they simply consist of local `qbits' defined on
the links of the lattice. Hence, in order to produce fermionic excitations
in the bulk, one must have bulk topological order. This is readily
seen as arising by `gauging' the fermion parity. That is, the microscopic
symmetry that is always present for physical fermion - the conservation
of their number modulo 2 - here is attributed to their $\mathbb{Z}_{2}$
gauge charge. Thus one has an emergent $\mathbb{Z}_{2}$ gauge theory
in the bulk. Additionally, there are $\mathbb{Z}_{2}$ flux loops
in the bulk that are gapped. Fermions circling these flux lines pick
up a $\pi$ phase shift - indicating that the fermion parity has been
gauged. However this bosonic theory is readily related to the free
fermion topological phases by the following slave particle construction:
say we decompose the physical boson destruction operator into a pair
of fermionic partons, $b_{r}=f_{r\uparrow}f_{r\downarrow}$. Time
reversal is assumed to act projectively on the fermions $f_{\uparrow}\overset{\mathcal{T}}{\rightarrow}f_{\downarrow};\, f_{\downarrow}\overset{\mathcal{T}}{\rightarrow}-f_{\uparrow}$.
The fermions are then governed by a mean field Hamiltonian in a topological
phase: $H=\sum_{ij}\left(t_{ij}f_{i}^{\dagger}f_{j}+\Delta_{ij}f_{i}f_{j}+{\rm h.c.}\right)$
(here $i$ refers to both space and spin index). To `ungauge' the
$\mathbb{Z}_{2}$ symmetry one simply introduces fundamental fermions
$c_{r\sigma}$ in the model and condenses the pair amplitude $\langle\sum_{\sigma}c_{r\sigma}^{\dagger}f_{r\sigma}\rangle$.
Now, the $\mathbb{Z}_{2}$ flux loops are confined since they have
nontrivial mutual statistics with the condensate. Hence the bulk topological
order is removed, and one realizes the short range entangled topological
phase in the bulk.

\section{Abelian Surface Topological Orders and the sixteen fold way \label{sec:SemionFermion}}

\subsection{The Semion-Fermion State and Time reversal symmetry\label{sub:The-Semion-Fermion-State}}

So far, we have been discussing $SO(3)_{6}$, a non-abelian theory
which we propose to be a candidate for the ${\cal T}$-symmetric surface
termination of a $\nu=1\,({\rm mod}\,2)$ topological superconductor.
Now let us consider even $\nu$. In this section, we introduce an
Abelian fermionic theory, which we call the \emph{semion-fermion}
theory, which turns out to also have an anomalous realization of ${\cal T}$.
The theory is the product of $U(1)_{2}$, which describes the universality
class of bosonic fractional quantum Hall systems at $1/2$ filling
and has quasiparticle content $\{1,s\}$, with $s$ a semion ($\theta_{s}=i)$,
and the trivial fermionic theory, with quasiparticle content $\{1,f\}$,
$f$ being a fermion. Letting $\tilde{s}=sf$, we obtain the quasiparticles
$\{1,s,\tilde{s},f\}$ with topological spins $\{1,i,-i,-1\}$ (labeled
$Z_{2}^{\frac{1}{2}}\times Z_{2}^{1}$ in the notation of Ref. \citep{Parsa}).
Potentially, one could imagine that this theory may be time reversal
symmetric if ${\cal T}$ exchanged $s$ and $\tilde{s}$. However,
this is not possible in a $2d$ system. This time the simple chiral
central charge argument does not work. Instead, we argue as follows:

First, consider gauging the $\mathbb{Z}_{2}$ fermion parity symmetry;
when this is done, the resulting theory is \emph{bosonic}, and the
electron carries $\mathbb{Z}_{2}$ gauge charge. It is hence described
by a unitary modular tensor category, called a modular extension of
the original fermionic theory. Note that a modular extension is not
unique: indeed, the trivial theory $\{1,f\}$, which has no anyons,
has $16$ possible modular extensions \citep{Kitaev}. These result
from the fact that even a trivial theory can have non-trivial behavior
of $\pi$ fluxes - e.g. a $p+ip$ superconductor is trivial in that
it has no anyons, but its modular extension - the Ising theory $\{1,\sigma,f\}$
- is different from the modular extension $\{1,e,m,f\}$ of the trivial
2D fermionic state.

Now consider the sixteen modular extensions of $\{1,s,\tilde{s},f\}$,
obtained by taking the product of $(1,s)$ with Kitaev's sixteen modular
extensions of $\{1,f\}$; these have chiral central charges $j/2\,({\rm mod}\,8)$,
$j=0,\ldots,15$, and $j=0$ could potentially be ${\cal T}$-invariant.
Since $U(1)_{2}$ has chiral central charge $1$, this $j=0$ theory
is the product of $U(1)_{2}$ and the $\nu=-2$ theory of \citep{Kitaev};
the latter has quasiparticles $\{1,a,\bar{a},f\}$ with $aa=\bar{a}\bar{a}=f$,
and $\theta_{a}=\theta_{\bar{a}}=e^{\pi i/4}$. Let us show that there
is no way that ${\cal T}$ could act and be consistent with the fusion
rules. First, note that ${\cal T}$ has to fix $1,f$ and exchange
$s,sf$. On the remaining quasiparticles it must do one of two things:
either $a\leftrightarrow sa$ and $\bar{a}\leftrightarrow s\bar{a}$,
or $\bar{a}\leftrightarrow sa$ and $a\leftrightarrow s\bar{a}$.
In the first case, $sf=a(sa)$ goes to $(sa)a=sf$ under ${\cal T}$,
which is inconsistent, and in the second case it also goes to $(s\bar{a})\bar{a}=sf$,
again an inconsistency.

This argument shows that the original fermionic $\{1,s,\tilde{s},f\}$
theory cannot exist in a 2D ${\cal T}$-symmetric system, assuming
that the above theories exhaust the modular extensions of $\{1,s,\tilde{s},f\}$.
The following argument establishes that this is so: given any realization
of $\{1,s,\tilde{s},f\}$, and forgetting ${\cal T}$, we can continuously
deform the Hamiltonian to any other one in this universality class.
In particular, we can choose a special one, which is effectively a
bosonic fractional quantum Hall layer in parallel with some $p+ip$
superconductor; we know that gauging this will yield one of our 16
modular extensions. Since the type of modular extension one obtains
after gauging is a discrete object, it cannot change under continuous
deformation. Hence, gauging any fermionic realization of $\{1,s,\tilde{s},f\}$
- including the putative ${\cal T}$-invariant one we started with
- must yield one of the 16 above theories, none of which are ${\cal {\cal T}}$-invariant.
Hence we have reached a contradiction.

However, it is possible to realize this theory in a ${\cal T}$ symmetric
way on the surface of a 3D topological phase. Just as for the case
of $SO(3)_{6}$, we can define a ${\cal T}$symmetry that exchanges
$s\leftrightarrow\tilde{s}$ and acts with appropriate phase factors
on the fusion vertices. We once again find that demanding ${\cal T}$
to commute with $F$ and $R$ symbols requires ${\cal T}^{2}=-1$
on the $(s,\tilde{s},f)$ vertices.

\subsubsection*{Walker-Wang model:}

Once again, we can build a Walker-Wang model based on the semion-fermion
theory. Because the theory is Abelian, we can actually use a simplified
cubic lattice instead of the original trivalent one here \citep{Burnell}.
The definition of ${\cal T}$-symmetry also follows as in the $SO(3)_{6}$
case: Haldane chains are bound to the $f$ lines, so that $f$ becomes
a deconfined bulk fermion with ${\cal T}^{2}=-1$: the electron. The
Walker-Wang model once again describes the $\mathbb{Z}_{2}$-gauged
version of the topological superconductor. For completeness, note
that the $F$ and $R$ symbols of the semion-fermion theory are just
products of those of the $\{1,f\}$ theory, in which the only non-trivial
symbol is $R_{1}^{f,f}=-1$, and the $\{1,s\}$ theory, defined by
$R_{1}^{s,s}=i$ and $[F_{s}^{s,s,s}]_{1,1}=-1$.

\subsubsection*{Coupled Layer construction of the Semion-Fermion Surface Topological
Order:}

We provide a coupled layer construction of a 3D topological superconductor
with a surface termination that realizes the semion-fermion Abelian
topological order $\{1,f\}\times U(1)_{2}=\left\{ 1,\, s,\tilde{\, s},\: f\right\} $
with time reversal symmetry. The building blocks are (i) a $\mathbb{Z}_{2}$
toric code topological order with 4 quasiparticles, $\{1,\, e,\, m,\,\psi\}$,
in which time reversal exchanges the two bosonic particles . This
is only possible if the fermion transforms projectively under time
reversal symmetry as discussed in the Appendix, Sec.\ref{sec:AppendixA}.
(ii) a doubled semion topological order also with 4 quasiparticles,
$\left\{ 1,\, S,\, S',\: b\right\} $ where time reversal exchanges
the semion and anti-semion. Collecting together the action of time
reversal on these excitations:

\begin{eqnarray}
e\overset{\mathcal{T}}{\leftrightarrow}m, & S\overset{\mathcal{T}}{\leftrightarrow}S', & \psi\overset{\mathcal{T}^{2}}{\leftrightarrow}-\psi\label{eq:Trev}
\end{eqnarray}

\begin{figure}
\includegraphics[scale=0.35]{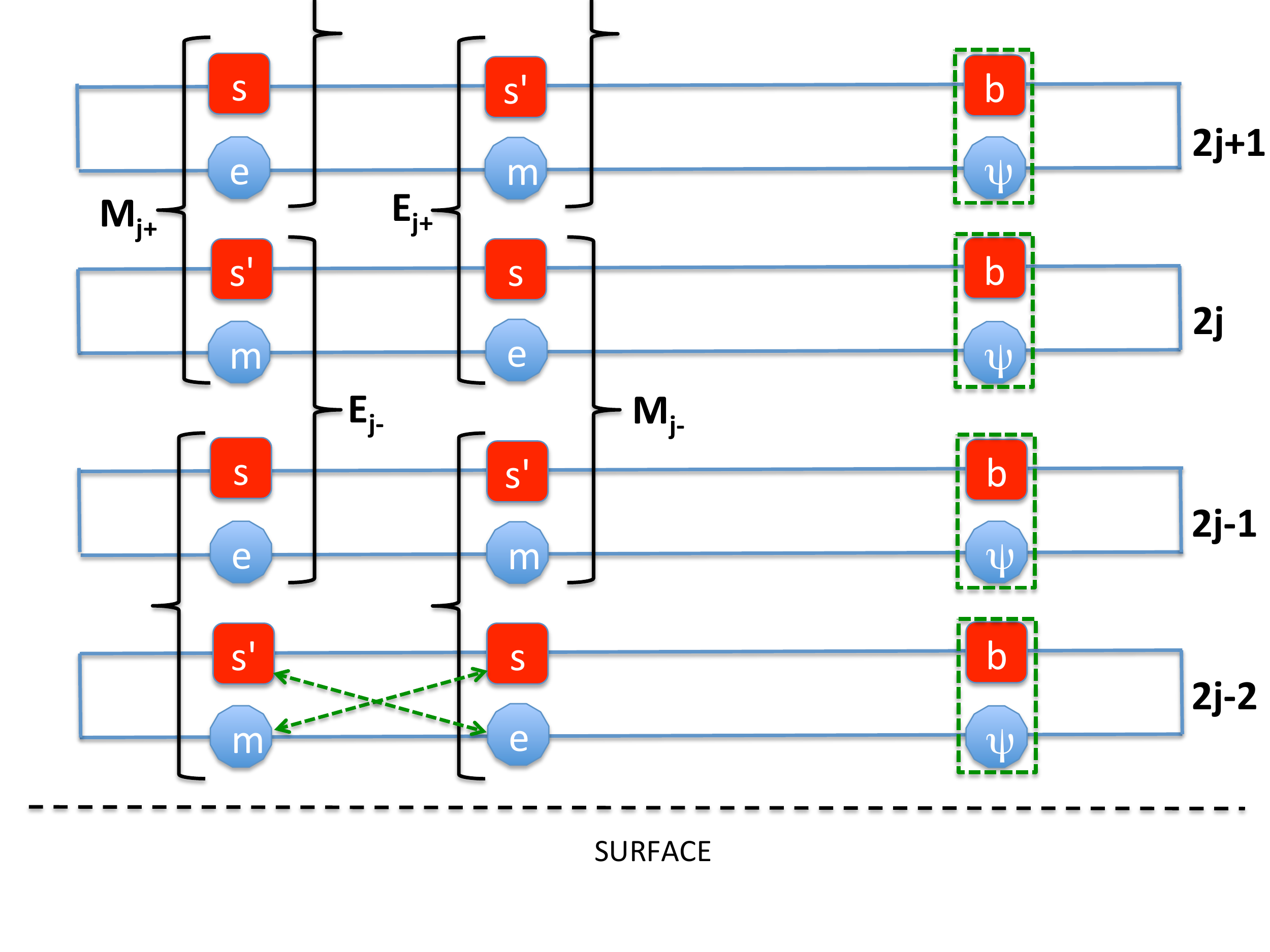}\caption{\label{Fig:CoupledLayer}Coupled layer construction of semion-fermion
topological order on the surface of a 3D topological superconductor
(with even $\nu$). Each layer has a double-semion model and a ${\cal Z}_{2}$
gauge theory. The composite excitations enclosed by the brackets can
be condensed simultaneously after which the only deconfined excitations
in the bulk are electrons $c=b\psi$ shown by the dashed green boxes.
At the surface, two additional excitations emerge - shown by the dashed
green lines -which are exchanged by ${\cal T}$. A key ingredient
is that the two bosons (electric and magnetic charges) of the ${\cal Z}_{2}$
gauge theory are exchanged by ${\cal T}$ which requires ${\cal T}^{2}=-1$
acting on electrons.}
\end{figure}
Consider a layered structure where in each layer, a topological state
of each of these two types is present (see Figure \ref{Fig:CoupledLayer}).
We define a unit cell to consist of pairs of these layers, and unit
cell $j$ will be consisting of layers $2j$ and $2j+1$. Now consider
the four bosonic operators associated with every unit cell (we suppress
the coordinates that represent the spatial position in the plane):

\begin{equation}
E_{j+}=e_{2j}S_{2j}m_{2j+1}S'_{2j+1}
\end{equation}
\begin{equation}
M_{j+}=m_{2j}S'_{2j}e_{2j+1}S{}_{2j+1}
\end{equation}
\begin{equation}
E_{j-}=e_{2j}S_{2j}m_{2j-1}S'_{2j-1}
\end{equation}
\begin{equation}
M_{j-}=m_{2j}S'_{2j}e_{2j-1}S{}_{2j-1}
\end{equation}

It is readily seen that this set of four operators are bosonic and
commute with one another. Hence they can be simultaneously `condensed'.
The resulting state will confine any excitation that has nontrivial
mutual statistics with these objects. In a system with periodic boundary
conditions, the only excitation that has trivial mutual statistics
with these four condensates is the bound state $f_{k}=\psi_{k}b_{k}$
in each layer. This is the deconfined fermion in the bulk of the system.
It is readily verified that all other anyon excitations are confined
in the bulk. 

If however a surface is present, one can identify anyonic excitations
that are deconfined within the surface. Consider a semi-infinite system
with layers indexed by $j=0,\,1,\,\dots\infty$. Therefore, the condensates
$E_{j=0-},\: M_{j=0-}$ are absent. Then, the surface excitations
that are now liberated (Fig:\ref{Fig:CoupledLayer}) are $s_{0}=e_{0}S'_{0},$
and $\tilde{s}=m{}_{0}S{}_{0}$. Their combination $s_{0}\tilde{s}_{0}=e_{0}s'_{0}m_{0}s_{0}=\psi_{0}b_{0}=f_{0}$
is just the fermion in the theory. Also, note using Eqn. \ref{eq:Trev},
that under time reversal symmetry:
\begin{equation}
s_{0}\overset{\mathcal{T}}{\leftrightarrow}\tilde{s_{0}}
\end{equation}

This is exactly the surface topological order that we sought. Moreover
it transforms under time reversal in exactly the way we require.

We note a couple of important points that have been alluded to while
discussing the alternate Walker-Wang construction. First, as discussed
in the Appendix, the transformation law for fermions $\psi$ and hence
for $f$ require that ${\cal T}^{2}=-1$ when acting on these objects.
Second, we note that the models used here are entirely bosonic - i.e.
they do not contain microscopic fermion excitations with trivial statistics
with all other particles, that can be identified with electrons. Although
there are fermionic excitations $f$ that are deconfined in the bulk,
these are emergent fermions, and hence have mutual statistics with
a $\mathbb{Z}_{2}$ flux loop excitation. This leads to bulk $\mathbb{Z}_{2}$
topological order. To rectify this and produce a fermionic model free
of bulk topological order, we introduce electron operators $c_{\sigma}$
that transform as Kramers pairs under time reversal symmetry. Then,
one can condense the combination $c^{\dagger}f$ which is a boson
that transforms trivially under time reversal symmetry, since both
components are Kramers pairs. As a consequence, flux excitations in
the bulk of the system are confined since they have mutual statistics
with this condensate, and there is no remaining bulk topological order.
Now, the $f$ fermion may be interpreted as the physical electron.

\subsection{Connecting the Abelian Semion-Fermion State to free fermion TScs}

In this section we discuss a physical argument that links the Abelian
semion fermion state with the free fermion $\nu=2$ TSc, with a pair
of Majorana cones on the surface. Our key idea is to exploit a larger
symmetry that is allowed by this problem - a U(1) symmetry that emerges
on allowing for rotations between the fermions in the pair of Majorana
modes, to constrain the surface topological order. One may then constrain
a candidate surface topological order by requiring it to satisfy a
number of physical criteria. The resulting topological order (that
respects the additional U(1) symmetry) that we conjecture is the same
as the T-Pfaffian which was discussed in the context of Topological
Insulator (TI) surface topological order. However, here, there are
important differences from the case of TIs, since time reversal is
combined differently with charge conservation. While in the case of
TIs, the charge is left invariant under $\mathcal{T}$, here the charge
will change sign under time reversal. Therefore `charge' here should
be viewed more appropriately as a component of spin (e.g. $S_{z}$).
Technically, this means that for the case of the insulators, the symmetry
is $U(1)\rtimes\mathcal{T}$ while here it is $U(1)\times\mathcal{T}$.
On destroying the artificial U(1) symmetry by condensing an appropriate
excitation, we are left with the the Abelian semion-fermion model,
which provides evidence for our assertion of the connection with the
$\nu=2$ TSc. A more systematic derivation of the connection to free
fermion TScs will appear in a forthcoming publication. 

Consider the free fermion surface state of a $\nu=2$ topological
superconductor: 
\begin{equation}
{\mathcal{H}}=\sum_{a=1}^{2}\chi_{a}^{T}\left(p_{x}\sigma_{x}+p_{y}\sigma_{z}\right)\chi_{a}\label{H2}
\end{equation}
Here $\chi_{a}^{T}=\left(\chi_{a}^{\uparrow},\,\chi_{a}^{\downarrow}\right)$
are two component Majorana fields and $\sigma$ are the Pauli matrices
in the usual representation, which act on the Majorana indices. The
flavor index $a=1,\,\dots\nu$. That is, there are $\nu$ `right handed'
Majorana cones on the 2D surface. 

The time reversal symmetry transformation is: 
\begin{equation}
\chi_{a}^{\uparrow}\xrightarrow{\mathcal{T}}\chi_{a}^{\downarrow}\,{\rm and}\,\chi_{a}^{\downarrow}\xrightarrow{\mathcal{T}}-\chi_{a}^{\uparrow}
\end{equation}
which leaves the Hamiltonian in Eqn. \ref{H2} invariant. The $O(2)=U(1)$
symmetry is obtained by rotating between the two flavors, so the complex
fermions: 
\begin{equation}
\psi_{\sigma}=\chi_{1}^{\sigma}-i\chi_{2}^{\sigma}
\end{equation}
transform as: $\psi_{\sigma}\rightarrow e^{i\varphi}\psi_{\sigma}$
under the U(1) rotation. In these variables, the Hamiltonian \ref{H2}
is:
\begin{equation}
{\mathcal{H}}=\psi^{\dagger}\left(p_{x}\sigma_{x}+p_{y}\sigma_{z}\right)\psi
\end{equation}
where the Pauli matrices act on the spin $\sigma$ indices, which
are suppressed. This is exactly the dispersion of the Dirac surface
state of a Topological \emph{Insulator}. The important difference
though is the transformation under time reversal symmetry. Under time
reversal,

\begin{equation}
\psi_{\uparrow}\xrightarrow{\mathcal{T}}\psi_{\downarrow}^{\dagger}{\rm \: and}\:\psi_{\downarrow}\xrightarrow{\mathcal{T}}-\psi_{\uparrow}^{\dagger}.
\end{equation}
Thus the U(1) charge is reversed under $\mathcal{T}$. Moreover, the
Cooper pair operator, defined as $e^{i\Phi_{{\rm Cooper}}}=\psi_{\uparrow}^{\dag}\psi_{\downarrow}^{\dag}$,
transforms under time reversal symmetry as $\Phi_{{\rm Cooper}}\rightarrow\Phi_{{\rm Cooper}}+\pi$.
This implies that a Cooper pair condensate automatically breaks time
reversal symmetry in this case. Thus there are no gapping terms for
this surface at the quadratic level that preserve $\mathcal{T}$,
as expected for the TSc surface. 

Since with U(1) symmetry the $\nu=2$ TSc maps to the single Dirac
cone surface state, we can use our knowledge of the topological insulator
surface topological order (STO) to guess the relevant state. The charge
conserving STO, when realized purely in a 2D system must break $\mathcal{T}$
symmetry and have $\sigma_{xy}=1/2$ and $\kappa_{xy}/T=1/2$ as argued
for the TI surface state, by appropriately gapping the single Dirac
cone with a $\mathcal{T}$ breaking mass term. Furthermore, given
the presence of Majorana fermions in vortices on the surface when
superconductivity is induced, we expect Ising topological order to
be involved. The simplest such theory that has a chance of being $\mathcal{T}$
symmetric at the level of statistics is the T-Pfaffian\citep{ParsaTI,ChenTI},
a 12 particle topological order which is a subset of ${\rm Ising}\times U(1)_{-8}$,
where the latter is the Laughlin $\nu_{{\rm QH}}=-1/8$ of charge
2e Cooper pairs. The $U(1)_{-8}$ leads, in a 2d system with the same
topological order, to a Hall conductance of $\sigma_{xy}=-\frac{1}{2}\frac{e^{2}}{h}$
. Also, on combining with the neutral sector of the Ising theory,
a simple count of all edge modes reveals $\kappa_{xy}/T=-\frac{1}{2}$,
as required. The ${\rm Ising}\times U(1)_{-8}$, theory has 24 particles.
However, an electron like excitation with charge `e' is identified
and this particle is required to have trivial statistics with all
other particles, reducing the number of quasiparticles to 12. The
quasiparticles are labeled $X_{q}$ where $X\in\{I,\,\sigma,\,\psi\}$
is the Ising part and $q=0,\dots,\,7$ is the $U(1)_{-8}$ part, and
they carry a charge $2e\frac{q}{8}$. The electron is identified as
$\psi_{4}$, and locality of this excitation restricts $q$ to be
even in $\{I_{q},\,\psi_{q}\}$ particles and $q$ odd in the $\{\sigma_{q}\}$.
Note, here the charge is reversed under $\mathcal{T}$, i.e. $q\rightarrow-q$
under $\mathcal{T}$. 

\begin{table}
\begin{tabular}{|c|c|c|c|c|c|c|c|c|}
\hline 
$X\downarrow k\rightarrow$ & 0 & 1 & 2 & 3 & 4 & 5 & 6 & 7\tabularnewline
\hline 
\hline 
I & +1 &  & +i &  & +1 &  & +i & \tabularnewline
\hline 
$\sigma$ &  & 1 &  & -1 &  & -1 &  & 1\tabularnewline
\hline 
$\psi$ & -1 &  & -i &  & -1 &  & -i & \tabularnewline
\hline 
${\rm {\color{blue}Charge}}$ & 0 & e/4 & e/2 & 3e/4 & e & 5e/4 & 3e/2 & 7e/4\tabularnewline
\hline 
\end{tabular}

\caption{Topological spins and charges of the T-Pfaffian State with $U(1)\times\mathcal{T}$
symmetry. Time-reversal reverses the charge $q\rightarrow-q$ (mod
8). This state is compatible with a surface electronic structure which
is a Dirac cone (or $\nu=2$ Majorana cones). The electron is identified
as $\psi_{4}$ and has charge `e' and $\mathcal{T}^{2}=-1$. On condensing
the boson $I_{4}$ (which breaks U(1) symmetry since it carries charge
`e' but preserves $\mathcal{T}$ since $\mathcal{T}^{2}=+1$) one
obtains the Abelian semion-fermion state, the proposed topological
order for the surface of a $\nu=2$ TSc. }
\end{table}

The excitations left invariant by time reversal symmetry then are
$I_{0},\,\psi_{0},\, I_{4},\,\psi_{4}$, which can be assigned$\mathcal{T}^{2}$
values. In addition, $I_{2},\,\psi_{2}$ are mapped to $\psi_{6},\, I_{6}$
under $\mathcal{T}$, which differ from the first pair by an electron,
which is a local excitation, allowing for assignment of $\mathcal{T}^{2}$
value to the pair. Indeed, ${\cal T}$ is fermionic when acting on
the pair $(I_{2,}\psi_{6})$, and therefore ${\cal {\cal T}}^{2}=\pm i(-1)^{F}$;
details of this projective representation of fermion parity and time
reversal symmetries are worked out in detail in the next section,
in the context of the semion-fermion model. Finally, since $\sigma_{k}\rightarrow\sigma_{-k}$,
no assignment of $\mathcal{T}^{2}$ values is possible for them.

Note that $\psi_{0}$ has $\mathcal{T}^{2}=-1$ because it is a bound
state of $I_{2}$ and $\psi_{6}$, which are mutual semions that are
exchanged by $\mathcal{T}$ 
\footnote{A pair of particles that are interchanged by $\mathcal{T}$and have
mutual statistics $\eta=\pm$ 1 will, when fused together, carry $\mathcal{T}^{2}=\eta$.
This may be understood by regarding the action of $\mathcal{T}^{2}$
as taking one particle around the other.}. On the other hand, since $I_{4}$ is a bound state of a pair of
$I_{2}$ (or $\psi_{2}$), it will have $\mathcal{T}^{2}=+1$ (${\cal T}^{2}=({\cal T}_{1}{\cal T}_{2})({\cal T}_{1}{\cal T}_{2})=-{\cal T}_{1}^{2}{\cal T}_{2}^{2}=-(i)(i)=1$,
where ${\cal T}_{1/2}$ are the fermionic local actions of ${\cal T}$
near the two copies of $I_{2}$. This is elaborated in the following
section.). Thus $I_{4}$ is a charge $e$ boson with $\mathcal{T}^{2}=+1$.
Since $\psi_{4}$ differs from $I_{4}$ by $\psi_{0},$ it must have
$\mathcal{T}^{2}=-1$, consistent with its identification as an electron.

We can now consider breaking the U(1) charge conservation to obtain
a simpler STO for $\nu=2$ TSc. Consider condensing $I_{4}$. The
consequence is that the $\sigma_{k}$ are confined. The remaining
theory contains $\psi_{2\tilde{k}},\, I_{2\tilde{k}}$ where $\tilde{k}=0,\,1$.
(the particles with $\tilde{k}=2,\,3$ are obtained from these by
combining with $I_{4}$, which is condensed). This is just the Abelian
semion-fermion theory $\{1,\, s\}\{1,\, f\}$ discussed above. The
two possibilities $\nu=\pm2$ are naturally associated with the $\mathcal{T}^{2}$
assignments of the semion (which is possible since it differes from
the antisemion by a local excitation, the electron - see the next
section). Here, those two assignments are related to the corresponding
transformation law for $I_{2}$. This argument links the surface of
a $\nu=\pm2$ TSc with the Abelian semion-fermion state.

\subsection{Remaining Abelian topological orders and the sixteen-fold way}

\begin{table*}
\begin{tabular}{|c|c|c|>{\centering}p{4cm}|}
\hline 
Label & Topological Order & $\mathcal{T}$ action on anyons & $\mathcal{T}^{2}$ action on anyons\tabularnewline
\hline 
\hline 
$\nu=1$ (mod 2) & $\{1,\,{\mathcal{S}},\,{\mathcal{S}}f,\, f\}$ & ${\mathcal{S}}\rightarrow{\mathcal{S}}f$ & \tabularnewline
\hline 
$\nu=\pm2$ (mod 8) & $\{1,\, s\}\{1,\, f\}$ & $s\rightarrow sf$ & $\mathcal{T}^{2}=\pm i$ acting on $s$; $\mathcal{T}^{2}=\mp i$
acting on $sf$ \tabularnewline
\hline 
$\nu=\pm4$ & $\{1,\, s_{1}\}\{1,\, s_{2}\}\{1,\, f\}$ & $s_{1,\,2}\rightarrow s_{1,\,2}f$ & $\mathcal{T}^{2}=\pm i$ acting on $s_{1}$, $s_{2}$; $\mathcal{T}^{2}=\mp i$
acting on $s_{1}f$, $s_{2}f$; $\mathcal{T}^{2}=+1$ acting on $s_{1}s_{2}$\tabularnewline
\hline 
$\nu=8$ & $\{1,\, F_{1},\, F_{2},\, F_{3}\}\{1,\, f\}$ & $F_{i}\rightarrow F_{i}$ & $\mathcal{T}^{2}=+1$ acting on $F_{i}$\tabularnewline
\hline 
\end{tabular}

\label{Table:Sixteenfoldway}\caption{Summary of Surface Topological Orders of TScs: The first row is the
non-Abelian $SO(3)_{6}$ topological order corresponding to the odd
$\nu$ TSc, where$\mathcal{S}$ to represents the nonabelian excitation
with semionic topological spin. The second row is the Abelian semion-fermion
theory which we have argued is related to the $\nu=\pm2$ TSc. The
third is two copies of the semion-fermion. Finally, the last row is
four copies of the semion-fermion theory, which corresponds (modulo
the electron) to the three fermion topological order which is also
the surface topological order of a purely bosonic topological phase
with a $\mathcal{Z}_{2}$ classification. Thus 8 copies of the semion-fermion
topological order are trivial, i.e. realizable in a purely two dimensional
system with $\mathcal{T}$, which implies that interactions unravel
the $\nu=16$ TSc connecting it to the trivial phase, consistent with
\citep{Kitaevunpublished}. The fermion $f$ transforms as $\mathcal{T}^{2}=-1$.}
\end{table*}

One subtlety that we have glossed over for the semion-fermion theory
is the value of ${\cal T}^{2}$ for $s$ and $\tilde{s}$. At first
one might think that there should not be a well defined value for
${\cal T}^{2}$ for these anyons, since they are not fixed under ${\cal T}$.
However, because they only differ by an electron, which is a local
(albeit fermionic) degree of freedom, we should really think of $s$
and $\tilde{s}$ as odd and even sectors of the same deconfined quasiparticle.
Then ${\cal T}^{2}=e^{i\phi}(-1)^{F}$ must hold locally for this
quasiparticle, where $\phi$ is some phase. It is easy to see that
since ${\cal T}$ anti-commutes with $(-1)^{F}$, we must have $\phi=\pm\pi$,
i.e. ${\cal T}^{2}=\pm i$ on $s$ and $\tilde{s}$. In other words,
the value of ${\cal T}^{2}$ is equal to the value of the topological
spin on $(s,\tilde{s})$ up to a sign $\eta=\pm1$ that does not depend
on whether we are acting on $s$ or $\tilde{s}$. Thus the two possibilities
$\eta=\pm1$ correspond to two distinct theories. Indeed, flipping
the theory corresponding to $\eta=1$ (i.e. taking the parity conjugate)
gives the theory corresponding to $\eta=-1$, since such a flip reverses
the topological spin but does not change the ${\cal T}^{2}$ eigenvalue.
The theories with $\eta=\pm1$ thus correspond to $\nu$ and $-\nu$
respectively, for some even integer $\nu$ (argued to be congruent
to $2$ mod $4$ in the next section).

Two\emph{ copies of Semion-Fermion Theory:} Consider now taking two
copies of the $\eta=1$ semion-fermion theory. The result is (two
copies of $U(1)_{2}$ times $\{1,f\}$) with time reversal again exchanging
$s_{1}\leftrightarrow s_{1}f$ and $s_{2}\leftrightarrow s_{2}f$,
where $s_{1}$ and $s_{2}$ are the two semions, both with the same
topological spin. Then $s_{1}s_{2}$ is a fermion with ${\cal T}^{2}=1$.
To see this, let ${\cal T}_{1}$ and ${\cal T}_{2}$ be the local
actions of time reversal near $s_{1}$ and $s_{2}$ respectively.
They are both fermionic operators, i.e. they anti-commute with fermion
parity and hence with themselves. Then ${\cal T}={\cal T}_{1}{\cal T}_{2}$
is a local action of time reversal on $s_{1}s_{2}$. We then have

\begin{center}
${\cal T}^{2}=({\cal T}_{1}{\cal T}_{2})({\cal T}_{1}{\cal T}_{2})=-{\cal T}_{1}^{2}{\cal T}_{2}^{2}=-(i)^{2}=1$
\par\end{center}

Thus this state has particles $\left\{ {1,\, f,\, s_{1},\, s_{2},\, s_{1}f,\, s_{2}f,\, B,\, F}\right\} $,
where the fermion $F=s_{1}s_{2}$ transforms as $\mathcal{T}^{2}=+1$
and the the boson $B=s_{1}s_{2}f$ transforms as $\mathcal{T}^{2}=-1$.

\noindent Had we taken a copy of $\eta=1$ with a copy of $\eta=-1$,
$s_{1}s_{2}$ would have been a fermion with ${\cal T}^{2}=-1$ and
we would have been able to condense $s_{1}s_{2}f$ without breaking
${\cal T}$, and get rid of all the topological order, as expected.
However, in the present case we cannot do this, and in fact the double
semion-fermion theory cannot be realized in 2D with ${\cal T}$ symmetry.
Based on the arguments above we identify this topological order with
the surface of a TSc with $\nu=\pm4$.

\emph{Three fermion topological order:} Although we could argue the
nontrivialness of the double semion-fermion theory directly, we proceed
by taking two copies of this theory, and showing that the result cannot
be realized in 2D with ${\cal T}$ symmetry. This is a stronger result
that in particular implies that the double semion fermion theory cannot
be realized in 2D with ${\cal T}$ either. In this  theory, $s_{1}s_{2}s_{1}'s_{2}'$
is a ${\cal T}^{2}=1$ boson and can be condensed. The non-trivial
particles in the resulting theory are the bilinears $\{ss_{2}=s_{1}'s_{2}',s_{1}s_{1}'=s_{2}s_{2}',s_{1}s_{2}'=s_{2}s_{1}'\}$,
times $\{1,f\}$. The bilinears are all fermions and mutual semions,
which shows that this is the three fermion $\mathbb{Z}_{2}$ topological
order, times $\{1,f\}$. We can label the bilinears $\{e,m,\varepsilon\}$;
the fusion and associativity rules are then the same as the toric
code. Also, ${\cal T}^{2}=1$ for $e,m,\varepsilon$. It was argued
in \citep{bosWW} that the three fermion theory can be realized at
the surface of an interacting bosonic SPT with ${\cal T}$ symmetry.
In fact, we can argue that its fermionized version also corresponds
to a non-trivial ${\cal T}$-invariant topological superconductor,
as follows:

Supposing for a contradiction that the fermionized version could be
realized in 2D in a ${\cal T}$-invariant way, we could gauge fermion
parity symmetry to end up with a modular ${\cal T}$-invariant theory.
This would have to be the product of $\{1,e,m,\varepsilon\}$ with
a modular extension of $\{1,f\}$; because the result is assumed ${\cal T}$-invariant,
it must be non-chiral, forcing the modular extension of $\{1,f\}$
to be another copy $\{1,e',m',f\}$ of the three-fermion $\mathbb{Z}_{2}$
topological order. Since ${\cal T}^{2}=-1$ on $f$, we must have
${\cal T}^{2}=-1$ on one of the other non-trivial particles - say
$e'$ - while ${\cal T}^{2}=1$ on $m'$. But then we could condense
the ${\cal T}^{2}=1$ boson $mm'$ while maintaining ${\cal T}$-invariance.
The particles that remain after this condensation are ${\cal T}^{2}=-1$
self-bosons $E\equiv ee'=\varepsilon f,M\equiv ef=\varepsilon e'$,
and the ${\cal T}^{2}=1$ fermion $\Sigma\equiv m=m'$. But this is
just the toric code $\{1,E,M,\Sigma\}$ with the $E$ and $M$ bosons
being Kramers doublets. This was argued to be impossible to realize
in a time reversal invariant way purely in 2D in \citep{VandS}, which
argued that instead this topological order should be realized as the
surface of the within-cohomology bosonic ${\cal T}$ SPT (see Appendix
\ref{sec:Stability} for more details). Hence the $\{1,e,m,\varepsilon\}$
three fermion theory, times $\{1,f\}$, cannot be realized in a 2D
fermionic system with ${\cal T}$ symmetry.

\emph{The sixteen fold way:} However, two identical copies of the
``three fermion'' times $\{1,f\}$ theory can be condensed into
the trivial phase without breaking ${\cal T}$. We have argued that
the semion-fermion theory corresponds to $\nu=2$ mod $4$, which
puts the three fermion theory at $\nu=8$ mod 16, and shows that $\nu=0$
mod $16$ should be trivial, and hence correspond to a trivial topological
superconductor. This is consistent with the recent result of Kitaev,
showing that the free fermion classification of 3D topological superconductors
breaks down to $\mathbb{Z}_{16}$ \citep{Kitaevunpublished}. Combining
the Abelian topological orders we have constructed in this section
with $SO(3)_{6}$ allows us to construct a gapped ${\cal T}$-symmetric
surface termination for any free fermion topological superconductor.
These results are summarized in Table \ref{Table:Sixteenfoldway}

\section{Conclusion}

We have provided several examples of time reversal symmetric topological
orders, realized on the surface of 3D gapped fermionic systems, that
are impossible to realize in a purely 2D system with time reversal
symmetry. This immediately implies that these phases are 3D fermionic
SPT phases protected by ${\cal T}$. In particular, it is impossible
to confine the surface states without breaking symmetry or closing
an energy gap. If it were possible to do so, then one can eliminate
the surface state on one face of a slab, and be left with the `impossible'
2D topological order on the other. Since the slab can then be deformed
into a 2D system, this leads to a contradiction. It is natural to
identify these phases with the free fermion topological superconductors
protected by ${\cal T}$. However,we note that we have not proved
this equivalence, \emph{e.g.} by demonstrating that the free fermion
Majorana cone edge states can annihilate one of our surface topological
orders. If this could be shown, it would eliminate the logical possibility
that the phases we are describing are some yet unknown topological
superconductor of fermions which is only realized in the presence
of interactions. Potentially, a classification of interacting SPT
phases of fermions in 3D could resolve this question, but that is
also currently unavailable (Ref. \citep{Gu2012} attempts to study
this question in some other symmetry classes). These are important
directions for future work. 

However, there are several pieces of evidence that link the surface
topological orders we mention with the free fermion topological superconductors.
First, our topological orders are only possible if the electrons transform
as ${\cal T}^{2}=-1$. This is also a requirement to realize the class
DIII free fermion topological phases. Next, it is well known that
odd $\nu$ topological superconductors have an odd number of chiral
Majorana modes bound to surface domain walls between domains of opposite
${\cal T}$ breaking. Indeed this is what we would find for the $SO(3)_{6}$
surface if the topological order is removed while breaking ${\cal T}$.
Imagine placing a 2D realization of the same topological order on
the surface to obtain a quantum double model that can be confined
to completely destroy topological order. Of course, in this process
one breaks ${\cal T}$ symmetry by the choice of the 2D topological
order. If this is done in opposite ways, at the domain wall one expects
an odd number of Majorana modes since the edge central charge of $SO(3)_{6}$
is $c_{-}=\pm2\frac{1}{4}$. The difference between these two opposite
values is consistent with the odd number of Majorana modes at the
surface domain wall. Given the simple nature of the topological order
in this case, it is tempting to attribute this to $\nu=1$, although
a proof of this requires a way to bridge these different descriptions.
Similarly, if the semion-fermion topological order describes a 3D
free fermion topological superconductor, it must correspond to even
$\nu$, since it is compatible with a gapped edge when realized in
2D. According to the arguments above, it corresponds to $\nu=2$ mod
$4$. Similarly the double semion-fermion theory corresponds to $\nu=4$
mod $8$, and the ``three fermion'' times $\{1,f\}$ theory to $\nu=8$
mod 16. Taken together, these would then generate topological terminations
for all possible 3D free fermion ${\cal T}$-invariant topological
superconductors.

One can also discuss analogous surface terminations for topological
insulators. Shortly after this work appeared, several papers proposing
such terminations were submitted to the arXiv \citep{MaxTI,WangTI,ParsaTI,ChenTI}.
In the future we hope to connect our work on topological surface terminations
with other non-perturbative definitions of the surface of fermionic
SPT phases. Finally, we wish to emphasize the remarkable fact that
in some cases fermionic SPT phase provides a guarantee of non-Abelian
topological order - if the surface of a $\nu=1$ topological superconductor
is found to be gapped and ${\cal T}$ symmetric, it must contain non-Abelian
excitations. 

\begin{comment}
\begin{itemize}
\item We don't know how to assign integers to our surface terminations -
recent Kitaev result, 16 fold way. in that case we have found 8. Would
be nice to match the two approaches. Requires direct comparison with
free fermion surface states. At least tells us the sequence terminates.
\item Provides a strong coupling definition of the phase - analogous to
Luttinger liquid for the edge of 2D.
\end{itemize}
In addition to being a new surface termination of a well known bulk
phase, there are advantages associated with defining the bulk phase
in terms of this particular surface termination. The topologically
ordered boundary theory preserves all symmetries, but is contrast
to the metallic surface states of the free fermion models, provides
a non-perturbative description of the essential physics. We expect
this line of though will be useful in future discussions of interaction
effects in 3D topological superconductors and insulators.
\end{comment}

\section*{Acknowledgments}

XC is supported by the Miller Institute for Basic Research in Science
at UC Berkeley. XC thanks discussions with A. Kitaev, Z. Wang, F.
Burnell, P. Bonderson, T. Senthil, C. Wang and A. Potter. AV thanks
A. Kitaev and P. Bonderson for insightful discussions, and is supported
by ARO MURI grant W911NF-12-0461. LF thanks Z. Wang, M.P. Fisher,
and M. Metlitski for useful discussions.

\section*{Appendix A: $\mathbb{Z}_{2}$ topological order in which time reversal
exchanges electric and magnetic charges.\label{sec:AppendixA}}

Consider the toric code model with the $\mathbb{Z}_{2}$ topological
order, leading to 4 particles $\{1,\, e,\, m,\,\psi\}$, where the
first particle is the identity,the next two are bosons (electric and
magnetic charges) and the last is a fermion. The last three particles
all have mutual semionic statistics with one another. Consider the
action of time reversal symmetry $\mathcal{T}$. The quasiparticles
can transform projectively, and we can choose a pair of them to transform
as $\mathcal{T}^{2}=-1$. Let one of these particles be the fermion
$\psi$. The $\psi$ fermions are gapped in this topological phase.
Consider modifying their dispersion so that they undergo a quantum
phase transition into a topological superconductor (class DIII) of
which there is a $\mathbb{Z}_{2}$ classification in 2D. Now, we can
investigate how the excitations transform under time reversal symmetry.
The fermions $\psi$ still transform as $\mathcal{T}^{2}=-1$. The
electric and magnetic particles can both be interpreted as $\pi$
flux for the fermions. If we choose the electric particle, we can
label the object that is obtained by attaching it to the fermion as
the magnetic particle i.e. we use the fusion rule $\psi\times e=m$.
In the language of the free fermion topological superconductor, if
we call the $\pi$ flux the electric particle, then the magnetic particle
has the opposite fermion parity. Now, it is readily seen that under
time reversal symmetry: $e\overset{\mathcal{T}}{\leftrightarrow}m$.
This is related to the fact that fermion parity and time reversal
anti commute\citep{Qi2009} when acting on a $\pi$ flux in this phase
$\left[\mathcal{T\,}(-1)^{{\rm N_{f}}}=-(-1)^{{\rm N_{f}}}\:\mathcal{T}\right]$.
This can be seen by considering the class DIII topological superconductor
as a tensor product of up spin and down spin electrons in a $p_{x}+ip_{y}$
and a $p_{x}-ip_{y}$ state respectively. Since time reversal exchanges
these two systems they remain invariant. But a $\pi$ flux inserted
through both superconductors will trap a pair of Majorana zero modes
$\gamma_{\uparrow,\downarrow}$. Under time reversal we have: $\gamma_{\uparrow}\overset{\mathcal{T}}{\leftrightarrow}\gamma_{\downarrow}$
and $\gamma_{\downarrow}\overset{\mathcal{T}}{\leftrightarrow}-\gamma_{\uparrow}$.
This ensures that the fermion parity $i\gamma_{\uparrow}\gamma_{\downarrow}$
for the vortex changes sign under time reversal symmetry. 

We note that although this symmetry transformation can be achieved
in 2D, it cannot be described by the K-matrix formulation of Abelian
topological orders, despite the underlying topological order being
an abelian one, specifically that of the $\mathbb{Z}_{2}$ toric code.
Indeed, in a recent classification of Symmetry Enhanced Topological
orders, using the K-matrix technique, this state was not produced\citep{LV}.
Here, time reversal symmetry exchanges the electric and magnetic particles,
which have mutual statistics. In general it can be shown that when
time reversal symmetry is implemented within the K-matrix, it \emph{cannot}
exchange anyons with \emph{nontrivial mutual statistics}.

The proof is as follows. Consider the K matrix of a time reversal
invariant state. This is an even dimensional $(2N\times2N)$ symmetric
matrix which can be diagonalized and brought into the form:

\begin{equation}
K=\sum_{\alpha=1}^{N}\lambda_{\alpha}\left[L_{\alpha}L_{\alpha}^{T}-R_{\alpha}R_{\alpha}^{T}\right]
\end{equation}

\noindent Since this is a symmetric matrix, the eigenvalues and eigenvectors
are real. The pairing of eigenvalues results from the fact that the
sign of the eigenvalues $\lambda$ refer to the chirality of edge
modes which should be coupled into time reversal symmetric nonchiral
pairs. Hence for every left mover there must be a right mover which
is related by time reversal $L_{\alpha}\overset{\mathcal{T}}{\leftrightarrow}R_{\alpha}$
. Now consider a quasiparticle represented by the integer vector $l$
and its time reversed partner $\tilde{l}$. Now expanding them in
terms of the eigenstates:

\begin{eqnarray*}
l & = & \sum_{\alpha=1}^{N}\left[a_{\alpha}L_{\alpha}+b_{\alpha}R_{\alpha}\right]
\end{eqnarray*}

\noindent where the coefficients are real numbers. Using the transformation
of eigenvectors under time reversal symmetry, we can write the time
reversed partner as:

\begin{eqnarray*}
\tilde{l} & = & \sum_{\alpha=1}^{N}\left[b_{\alpha}L_{\alpha}+a{}_{\alpha}R_{\alpha}\right]
\end{eqnarray*}

\noindent Now, it is readily verified that the two quasiparticles
have trivial mutual statistics since:

\begin{eqnarray*}
l^{T}K^{-1}\tilde{l} & = & 0
\end{eqnarray*}

\section*{\noindent Appendix B: Stability of 3D Bosonic SPT phases in the presence
of fermions\label{sec:Stability}}

Recently, symmetry protected topological phases of bosons were classified
- for example, with just time reversal ${\cal T}$ symmetry (the bosonic
analog of the class DIII systems discussed here), it was found that
there were three nontrivial phases composed together in a $\mathbb{Z}_{2}\times\mathbb{Z}_{2}$
structure. The nontrivial phases are generated by (i) a group cohomology
state and another state (ii) based on Kitaev's E$_{8}$ state that
lies outside the `group cohomology' classification.

When discussing bosonic SPT phases one assumes that the bosons are
fundamental particles. However, we could consider the possibility
that the bosons are composites (like spins or Cooper pairs) of fermions.
In this case there will also be gapped fermions in addition to the
bosonic degrees of freedom. Sometimes, the bosonic topological phase
can be unwound in the presence of fermions, or can be related to one
of the free fermion topological phases. Several instances of this
type were discussed in the context of interacting fermionic SPT phases
in 2D. Here we will discuss the 3D systems, with special reference
to the topological superconductors.

A useful tool to discuss this question in the 3D case, is the topologically
ordered surface termination of the 3D bosonic SPT phases. Based on
this we will argue that the $\mathbb{Z}_{2}\times\mathbb{Z}_{2}$
classification of bosonic SPTs is reduced to a single $\mathbb{Z}_{2}$
in the presence of fermions. Let us begin by discussing the surface
topological orders for the two Bosonic SPT states are (i) the three
fermion state, with quasiparticles $\{1,\, f_{1},\, f_{2},\, f_{3}\}$
where the three fermion have mutual semionic statistics and transform
with $T^{2}=+1$ under time reversal, and (ii) the $\mathbb{Z}_{2}$
toric code topological order $\{1,\, v_{1},v_{2},\, f\}$ where the
two bosonic excitations $v_{1},\, v_{2}$ both transform as $T^{2}=-1$
Kramers doublets.

We argue that this classification collapses to a single $\mathbb{Z}_{2}$
in the presence of fermions $\psi$ which transform as ${\cal T}^{2}=-1$.
To see this consider putting together this fundamental fermion $\{1,\,\psi\}$
with these topological orders. In each case we obtain:

\begin{table}[htdp]
\caption{The group cohomology state surface topological order with $v_{1,2}$
that transform as ${\cal T}^{2}=-1$ attached to electrons.The $\pm$
value is the action of ${\cal T}^{2}$ on that quasiparticle.}

\begin{centering}
\begin{tabular}{|c|c|c|c|c|}
\hline 
Label  & 0  & 1 & 2  & 3 \tabularnewline
\hline 
B  & \textbf{1} +  & $v_{1}$ $-$ & $v_{2}$ $-$ & $f\psi$ $-$\tabularnewline
\hline 
F  & $\psi$ $-$ & $v_{1}\psi$ + & $v_{2}\psi$ + & $f$ +\tabularnewline
\hline 
\end{tabular}
\par\end{centering}

\label{default} 
\end{table}

\begin{table}[htdp]
\caption{The beyond group cohomology state and related three fermion surface
topological order, attached to electrons. The $\pm$ value is the
action of ${\cal T}^{2}$ on that quasiparticle.}

\begin{centering}
\begin{tabular}{|c|c|c|c|c|}
\hline 
Label  & 0  & 1 & 2  & 3 \tabularnewline
\hline 
B  & \textbf{1} +  & $f_{1}\psi$ $-$ & $f_{2}\psi$ $-$ & $f_{3}\psi$ $-$\tabularnewline
\hline 
F  & $\psi$ $-$ & $f_{1}$ + & $f_{2}$ + & $f_{3}$ +\tabularnewline
\hline 
\end{tabular}
\par\end{centering}

\label{default-1} 
\end{table}

Clearly the two tables above are identical - we can represent the
particles as $B_{0}\dots B_{3},\, F_{0}\dots F_{3}$ and they have
identical self and mutual statistics and transformation laws under
time reversal. Therefore these topological orders are identical in
the presence of fundamental fermions which transform under time reversal
symmetry as ${\cal T}^{2}=-1$. That is, some of the bosonic SPT phases
`unwind' in the presence of fundamental fermions%
\footnote{Related results have been obtained by C. Wang, A. Potter and T. Senthil
(to appear) in the context of topological insulators.%
}.

\section{Appendix C: Hamiltonian for decorated Walker Wang model\label{sec:Appendix-C:-Hamiltonian}}

In this section we describe in detail the Hamiltonian for our exactly
solved model of an SPT realizing a gapped surface with $SO(3)_{6}$
topological order. As discussed in section \ref{sub:Caveat}, this
is a decorated bosonic Walker-Wang model, and is conjectured to describe
the universality class of the $\nu=1$ topological superconductor,
after $\mathbb{Z}_{2}$ fermion parity has been gauged. Although we
ultimately want such a bosonic model to be built out of Kramers singlets,
we first introduce auxilliary degrees of freedom which are Kramers
doublets - i.e. spin $1/2$'s, and write a Hamiltonian for this model.
As described in \ref{sub:Caveat}, there will be two spin $1/2$'s
per link, one associated with each endpoint vertex - together these
form a $4$ dimensional Kramers singlet Hilbert space ${\cal H}_{l}^{spin}$,
where $l$ labels the link. Thus the total Hilbert space associated
to each link is ${\cal H}_{l}^{WW}\otimes{\cal H}_{l}^{spin}$; the
Hilbert space of the model is the tensor product of these spaces over
all links.

The Hamiltonian has the general form

\begin{center}
$H=H_{const}+H_{plaq}.$
\par\end{center}

\noindent The first term is

\begin{center}
$H_{const}=V_{fusion}+V_{links}+V_{spins}+V_{vert}$
\par\end{center}

\noindent where $V_{fusion}$ is the fusion rule enforcing vertex
potential energy term present in the standard Walker Wang model, $V_{links}$
is a potential energy term preferring singlets along links not labeled
with $e$, $V_{spins}$ is an energetic preference for spin up or
down at a $(s,\tilde{s},e)$ vertex, depending on the sign of the
permutation that takes $(s,\tilde{s},e)$ into a counter-clockwise
labeling of the three links adjoining the vertex, and $V_{vert}$
is a term which acts on any vertex adjoined by exactly two $e$ lines
by forcing the spin $1/2$'s corresponding to the $e$'s into a singlet.
All of these terms will give an energy penalty of the same order $V$
to configurations that violate the constraints.

Below we will describe the plaquette terms $H_{plaq}$, which move
between different string net and spin configurations. These terms
will be modifications of the Walker Wang plaquette terms; in particular
they will be defined to act as $0$ on configurations which violate
any of the constraints in $H_{const}$ in a neighborhood of the plaquette
in question (i.e. on any of the links and vertices in Fig. \ref{fig:plaquette_term}).
Furthermore, they will commute with each other and with all the terms
in $H_{const}$. We will construct a state which satisfies all of
the constraints, and is the lowest eigenvalue state of all of the
plaquette terms - this will be our ground state, and will have the
form described in Section \ref{sub:Caveat}, as desired. Furthermore,
any other ground state must also satisfy all of the constraints, and
since the spin part of a configuration which satisfies all the $H_{const}$
constraints is uniquely determined by the anyon labels, this model
must have the same ground state degeneracy as the original Walker
Wang model - i.e. it is non-degenerate, according to arguments along
the lines of \citep{bosWW}.

It remains to construct $H_{plaq}$. We have

\begin{center}
$H_{plaq}=\sum_{P}\tilde{B}_{P}^{s},$
\par\end{center}

\noindent where the sum is over plaquettes $P$, and $\tilde{B}_{P}^{s}$
is an ``improved'' Walker-Wang plaquette term. To describe it, first
recall the original Walker-Wang plaquette term $B_{P}^{s}$ defined
in Eq. \ref{eq:WW_plaq}, which gives its matrix elements between
fusion rule respecting string net configurations (by definition it
is $0$ when acting on non-fusion rule respecting string net configurations).
Now, $\tilde{B}_{P}^{s}$ will by definition be $0$ unless all fusion
rules and all constraints in $H_{const}$ are satisfied; when they
are, we know that the spin part of the wavefunction is uniquely determined
by the string net labeling (we have to choose a fixed sign convention
for all of the spin singlets), so we can define $\tilde{B}_{P}^{s}$
to have nonzero matrix elements precisely between those states. The
operators $\tilde{B}_{P}^{s}$ so defined automatically commute with
all the constraints - because they act non-trivially only on configurations
that satisfy them, and move them to other configurations that satisfy
them - and they commute among each other for different plaquettes
$P$, for the same reason that the $B_{P}^{s}$ do. The ground state
of the original Walker-Wang model, with the unique spin configuration
slaved to it via the above constraints, is then also the lowest eigenvalue
state of all of the $\tilde{B}_{P}^{s}$, as desired.

\bibliographystyle{unsrtnat}
\bibliography{TopoSC}

\end{document}